\documentclass[11pt]{article}
\usepackage{amsmath,amssymb,color,graphics,epsfig}
\usepackage[compress]{cite}
\usepackage{hyperref}
\usepackage{ulem}

\usepackage{amsfonts}

\newcommand{\be}{\begin{equation}}
\newcommand{\ee}{\end{equation}}
\newcommand{\bea}{\setlength\arraycolsep{2pt} \begin{eqnarray}}
\newcommand{\eea}{\end{eqnarray}}
\newcommand{\nn}{\nonumber}

\def\ft#1#2{{\textstyle{\frac{\scriptstyle #1}{\scriptstyle #2} } }}

\def\0{{\sst{(0)}}}
\def\1{{\sst{(1)}}}
\def\2{{\sst{(2)}}}
\def\3{{\sst{(3)}}}
\def\4{{\sst{(4)}}}
\def\5{{\sst{(5)}}}
\def\6{{\sst{(6)}}}
\def\7{{\sst{(7)}}}
\def\8{{\sst{(8)}}}
\def\sst#1{{\scriptscriptstyle #1}}

\begin{document}

	
	\begin{center}
		{\Large {\bf Shadow images of compact objects in beyond Horndeski theory}}

		\vspace{20pt}
		
	{\large{Hyat Huang$^{1,2}$, Jutta Kunz$^2$ and Deeshani Mitra$^3$ }}

\vspace{10pt}

{\it $^1$College of Physics and Communication Electronics, Jiangxi Normal University, Nanchang 330022, China\\
$^2$  Institute of Physics, University of Oldenburg, Postfach 2503, D-26111 Oldenburg,
Germany\\
$^3$ Department of Physics, St. Xavier's College (Autonomous), Kolkata, India
}

		\vspace{40pt}
		
		\underline{ABSTRACT}
	\end{center}
	
A  beyond Horndeski theory is considered that admits wormholes, black holes and naked singularities. In this theory the shadow images of the black holes and the exotic compact objects (ECOs), illuminated by an optically and geometrically thin disk, are investigated.
The results show that the three kinds of objects cast unlike shadow images, in particular, because the different objects possess a different number of light rings. 
The different boundaries of the accretion disk also affect the images. 
This may provide further insight into the nature of the shadow images of massive compact objects.

\vfill {\footnotesize 
~\\
hyat@mail.bnu.edu.cn\\
jutta.kunz@uni-oldenburg.de
}\ \ \ \

\thispagestyle{empty}
	
	\pagebreak
\addtocontents{toc}{\protect\setcounter{tocdepth}{2}}
	

 
\section{Introduction}

Viable alternative theories of gravity are required to satisfy all known observational constraints.
First of all, for weak gravitational fields they should conform to the stringent solar system constraints \cite{Will:2005va,Will:2018bme}.
For strong gravitational fields, as present in the vicinity of compact objects like black holes or neutron stars, however, observations still leave a lot of freedom \cite{Berti:2015itd,Barack:2018yly,CANTATA:2021ktz}.

Recent observations of gravitational waves from the merger of black holes and neutron stars by the LIGO/VIRGO consortium have given us new insights into the strong gravity regime \cite{LIGOScientific:2016aoc,LIGOScientific:2017vwq}.
This also holds for the recent images of the black holes at the center of M87 and the Milky Way obtained by the Event Horizon Collaboration \cite{EventHorizonTelescope:2019dse,EventHorizonTelescope:2022wkp}.
On the other hand, pulsar observations have already put severe constraints on certain alternative theories (see e.g.~\cite{Shao:2017gwu,Freire:2022wcz}). All this experimental progress has triggered a great deal of renewed interest in the theoretical study of gravity.

General Relativity is a most successful gravitational theory, which predicts the existence of black holes. 
Furthermore, General Relativity has also inspired the study of exotic compact objects, like wormholes, boson stars, naked singularities and so on. 
However, the violation of
the energy conditions present in some of these objects may nearly rule them out in General Relativity. 
For example, a classical traversable wormhole in General Relativity needs exotic matter,
\cite{Morris:1988cz,Morris:1988tu}. 
Nevertheless, numerous studies consider 
that exotic compact objects might exist see e.g. \cite{Cardoso:2016oxy,Mark:2017dnq,Ou:2021efv,Cunha:2022gde}),
motivating our study of compact objects in alternative theories of gravity.

From a theoretical perspective a class of theories known as Horndeski theories has received much attention in recent years \cite{Horndeski:1974wa,Deffayet:2013lga}.
Horndeski theories involve additionally only a single scalar degree of freedom, lead to second order field equations, and are free of ghosts, which makes them rather attractive.
Moreover, they possess a set of free functions, thus yielding a variety of compact objects with distinct and interesting properties.

This is seen, for instance, in the subclass of Einstein-scalar-Gauss-Bonnet theories, where various types of hairy black holes arise \cite{Kanti:1995vq,Lu:2020iav,Doneva:2022ewd}, but also wormholes without the necessity for the presence of exotic matter \cite{Kanti:2011jz,Antoniou:2019awm}, as well as particle-like solutions \cite{Kleihaus:2019rbg}.
While most of these solutions are known only numerically, the black holes obtained by taking the $D \to 4$ limit of higher-dimensional Einstein-Gauss-Bonnet solutions \cite{Glavan:2019inb} and subsequently amended with a scalar field \cite{Lu:2020iav,Hennigar:2020lsl,Fernandes:2020nbq} are known in closed form.

Interestingly, Horndeski theories can be extended to obtain a class of theories referred to as beyond Horndeski theories \cite{Gleyzes:2014dya,Langlois:2015cwa,Crisostomi:2016czh}.
While the field equations of these theories possess higher order derivatives, the equations governing the propagating degrees of freedom are still of second order.
In addition, the theories retain the property of being free of ghosts.
Moreover, with so-called generalised disformal transformations, beyond Horndeski theories can be generated from Horndeski theories \cite{Gleyzes:2014qga,Crisostomi:2016tcp,Crisostomi:2016czh}.

On the other hand, theoretical investigations of shadow images of compact objects are quite fascinating. 
This applies not only to black holes in General Relativity \cite{Bambi:2012tg,Johannsen:2013vgc,Cunha:2015yba,Cunha:2016bpi,Gralla:2019xty,Younsi:2021dxe,Promsiri:2023rez} and in alternative gravities \cite{Zeng:2020dco,Peng:2020wun,Guo:2020zmf,Kumar:2020owy,Konoplya:2020bxa,Gyulchev:2021dvt,Zeng:2022fdm,Ye:2023qks}, but also to exotic compact objects like wormholes \cite{Bambi:2013nla,Schee:2021pdt,Bambi:2021qfo,Guerrero:2021pxt,Rahaman:2021web,Tsukamoto:2021fpp,Peng:2021osd,Guerrero:2022qkh,Delijski:2022jjj,Huang:2023yqd,Ishkaeva:2023xny}, naked singlarities \cite{Tavlayan:2023vbv,Vagnozzi:2022moj,Gyulchev:2020cvo,Joshi:2020tlq,Dey:2020bgo,Gyulchev:2019tvk,Shaikh:2018lcc,Deliyski:2023gik} or boson stars \cite{Rosa:2022toh,Rosa:2022tfv}. 
Previous studies have shown
that some of these exotic compact objects can mimic black holes with respect to their shadow images if their light ring properties are similar. 
But there are also studies showing that exotic compact objects may possess very special shadow images \cite{Huang:2023yqd,Rosa:2022toh,Rosa:2022tfv,Guerrero:2022qkh}.
One intriguing direction here is to examine the shadow images of different types of compact objects, like black holes, wormholes and naked singularities in a unified framework of a particular gravitational theory, 
providing a comparative study in this field.

Recently Bakopoulos et al.~\cite{Bakopoulos:2021liw} employed a disformal transformation to the (amended) Einstein-scalar-Gauss-Bonnet theory \cite{Lu:2020iav,Hennigar:2020lsl,Fernandes:2020nbq}, to obtain a family of beyond Horndeski theories.
Their particular interest focused on the disformally transformed black hole solution \cite{Lu:2020iav}, showing that by making appropriate parameter choices regular wormhole solutions arise \cite{Bakopoulos:2021liw}.
These beyond Horndeski theories feature black holes, wormholes and naked singularities.

The existence of all three types of solutions in closed form invites their further study and the comparison of their properties in the light of potential observations.
We here therefore consider the shadow images of these beyond Horndeski black holes, wormholes and naked singularities.
In section 2 we discuss the theoretical set-up and the three types of solutions.
We address the lightlike and timelike geodesics in these spacetimes in section 3 and discuss the images obtained in the presence of an accretion disk in section 4. We conclude in section 5.

\section{Set-up}\label{se1}

\subsection{Beyond Horndeski theory}

Horndeski theory features an additional degree of freedom. In shift-symmetric Horndeski theory the four functions $\{G_2, G_3, G_4, G_5\}$ are only functions of the kinetic term $X=-\ft{1}{2}\partial_\mu \phi\partial^\mu \phi$, where $\phi$ denotes the scalar field. 
Following \cite{Bakopoulos:2021liw}, we start from shift-symmetric Horndeski theory with action
\be
S_H=\int d^4 x \sqrt{-g}({\cal L}_2+{\cal L}_3+{\cal L}_4+{\cal L}_5),
\ee
where
\bea
{\cal L}_2&=& { G_{2}(X)}\\
{\cal L}_3&=&-{G_{3}(X)\square{\phi}}\nn\\
\mathcal{L}_4 &=& G_{4}(X)R + G_{4X}[(\square{\phi})^2 - \nabla_\mu \nabla_\nu \phi \nabla^\mu \nabla^\nu \phi]\nn\\
\mathcal{L}_5 &=& G_{5}(X)G_{\mu\nu}\nabla^\mu \nabla^\nu \phi - \frac{1}{6}G_{5X}[(\square{\phi})^3 - 3\square{\phi }\nabla^\mu \partial^\nu \phi \nabla_\mu \partial_\nu \phi + 2\nabla_\mu \partial_\nu \phi \nabla^\nu \partial^\rho \phi \nabla_\rho \partial^\mu \phi]
. \nn
\eea
To obtain the corresponding beyond Horndeski theory one needs to add the higher-order terms
\bea
\mathcal{L}^{bH}_{4} &=& F_{4}(X)\epsilon^{\mu\nu\rho\sigma} \epsilon^{\alpha\beta\gamma}_{\ \ \ \ \sigma}  \partial_{\mu}\phi\partial_{\alpha}\phi\nabla_{\nu}\partial_{\beta}\phi\nabla_{\rho}\partial_{\gamma}\phi\nn\\
\mathcal{L}^{bH}_{5} &=& F_{5}(X)\epsilon^{\mu\nu\rho\sigma}\epsilon^{\alpha\beta\gamma\delta}\partial_{\mu}\phi\partial_{\alpha}\phi\nabla_{\nu}\partial_{\beta}\phi\nabla_{\rho}\partial_{\gamma}\phi\nabla_{\sigma}\partial_{\delta}\phi ,
\eea
where, in order to avoid the Ostrogradski ghost degree of freedom, the following relation should hold \cite{BenAchour:2016fzp}
\be
XG_{5X}F_4=3F_5(G_4-2XG_{4X}),
\ee
and the subscript $X$ denotes the derivative with respect to $X$. 

By making use of disformal transformations one can start from a solution in Horndeski theory to obtain new solutions in beyond Horndeski theory \cite{Gleyzes:2014qga,Crisostomi:2016tcp,Crisostomi:2016czh}.
To distinguish quantities in Horndeski theory from the new transformed quantities in beyond Horndeski theory, we now denote the Horndeski quantities by barred quantities.
Starting from the Horndeski metric $\bar{g}_{\mu\nu}$ and employing a disformal transformation function $D(\bar X)$ then leads to the new beyond Horndeski metric $g_{\mu\nu}$
\be
g_{\mu\nu}=\bar{g}_{\mu\nu}-D(\bar{X})\nabla_\mu \bar{\phi}\nabla_{\nu}\bar{\phi}.
\ee

We next specify to static spherically symmetric solutions.
In this case the Horndeski metric can be parameterized by two functions of the radial coordinate $r$, $\bar h(r)$ and $\bar f(r)$,
\be
ds^2=-\bar{h} dt^2+\bar{f}^{-1}dr^2+r^2d\Omega^2_2 .
\ee
Note, that if the Horndeski scalar field $\bar \phi$ is also only a function of $r$, one obtains the transformed functions
\be
\phi=\bar{\phi},\quad h=\bar{h},\quad f=\ft{\bar{f}}{1+2D\bar{X}},\quad X=\ft{\bar{X}}{1+2D \bar{X}}.
\ee

\subsection{Beyond Horndeski solution}

Along with \cite{Lu:2020iav,Hennigar:2020lsl,Fernandes:2020nbq} we now specify the Horndeski theory by choosing the functions $G_i$, $i=2-5$, according to \cite{Bakopoulos:2021liw} (see also \cite{Kobayashi:2011nu})
\be
G_2=8\alpha X^2,\quad G_3=-8\alpha X,\quad G_4=1+4\alpha X, \quad G_5=-4 \alpha \log|X| ,
\ee
where $\alpha$ is the coupling constant of the theory.

The seed solution of the original Horndeski theory is given by \cite{Lu:2020iav}
\bea\label{ds}
&&\bar{h}=\bar{f}=1+\ft{r^2}{2\alpha}\bigg(1-\sqrt{1+\ft{8\alpha M}{r^3}}\bigg),\quad \bar{\phi}'=\ft{\sqrt{\bar{h}}-1}{r\sqrt{\bar{h}}},
\eea
with the prime denoting differentiation with respect to the radial coordinate.
For $\alpha \le M^2$ this metric
describes an asymptotically flat black hole with mass $M$. 

The beyond Horndeski solution is obtained by
the disformal transformation $D(\bar{X})$, leaving the metric function $\bar h$ and the scalar field $\bar \phi$ invariant and changing only the metric function $\bar f$, i.e.,
\be
h=\bar{h},\qquad f=\ft{\bar{h}}{1+2D(\bar{X})\bar{X}}=h W(\bar{X})^{-1},\qquad \phi=\bar{\phi},
\ee
where we have introduced the function $W(\bar{X})$, incorporating the disformal transformation \cite{Bakopoulos:2021liw} 
\be
W(\bar{X})=1+2D(\bar{X})\bar{X}.
\ee
Thus the line element becomes
\be\label{line1}
ds^2=-hdt^2+\ft{dr^2}{h W^{-1}}+r^2d\Omega^2_2.
\ee

Let us now specify the function $W(\bar{X})$, following again \cite{Bakopoulos:2021liw}.
While there is an infinite set of functions yielding new solutions, 
we now choose
\be
W^{-1}=1-\ft{r_0}{\lambda r}(1-\sqrt{h}) ,
\ee
with the parameters $r_0$ and $\lambda$.
The metric of the solution in the beyond Horndeski theory is then given in terms of the metric functions
\be\label{metricf}
h=1+\ft{r^2}{2\alpha}\bigg(1-\sqrt{1+\ft{8\alpha M}{r^3}}\bigg), \qquad f=h\bigg(1-\ft{r_0}{\lambda r}(1-\sqrt{h})\bigg).
\ee
This choice, in fact, allows us to describe three different types of compact objects, black holes, wormholes and naked singularities, as discussed below.

\subsection{Wormholes, black holes and naked singularities}

The solution \eqref{metricf} of the beyond Horndeski theory contains four parameters, namely $(\alpha, \lambda, M, r_0)$.
This allows for a lot of freedom in the kind of object it may describe, yielding 
wormholes, black holes or naked singularities depending on the choice of the parameters. 

The condition for obtaining black hole solutions is the presence of a horizon.
Suppose $r=r_h$ is a real solution of
\be
h(r_h)=0,
\ee
implying
\be
r_h=M \pm \sqrt{M^2-\alpha},
\ee
then $r_h=M + \sqrt{M^2-\alpha}$ is an outer black hole horizon, if $h(r)>0$ and $f(r)>0$ for $r>r_h$.
There is no real solution, if $\alpha\geq M^2$.

Wormhole solutions are obtained by requiring that $h(r_t)>0$ and $f(r_t)=0$ at the location $r=r_t$. At the same time, to make sure there is no black hole horizon outside $r>r_t$, the condition $h(r)>0$ should also hold for $r>r_t$. 
Then by solving
\be
W^{-1}(r_t)=0 ,
\ee
one obtains the location of a wormhole throat.
\footnote{In order to construct wormhole solutions easily, $r_t=r_0$ is introduced in Ref.~\cite{Bakopoulos:2021liw}, which leads to $h(r_0)=(1-\lambda)^2$ with $\lambda\in (0,1)$.
Then the four parameters $(r_0, M, \alpha, \lambda)$ must satisfy
$r_0 =\frac{M + {\sqrt{M^2 - \alpha  \lambda^{3}(2 - \lambda)^3} }}{\lambda(2 - \lambda)}$.
}

In other words, $r_t$ and $r_h$ are both real solutions for $\alpha\leq M^2, \lambda>0$. 
The condition for black holes, $r_t\leq r_h$, gives rise to 
\be
\ft{r_0(1-\sqrt{h(r_t)})}{\lambda}\leq M+\sqrt{M^2-\alpha}.
\ee
It then follows that the boundary surface is determined by $r_t=r_h$, which leads to  
\be
\ft{r_0}{\lambda}=M+\sqrt{M^2-\alpha}.
\ee

\begin{figure}[h]
\centering
\includegraphics[width=0.47\textwidth]{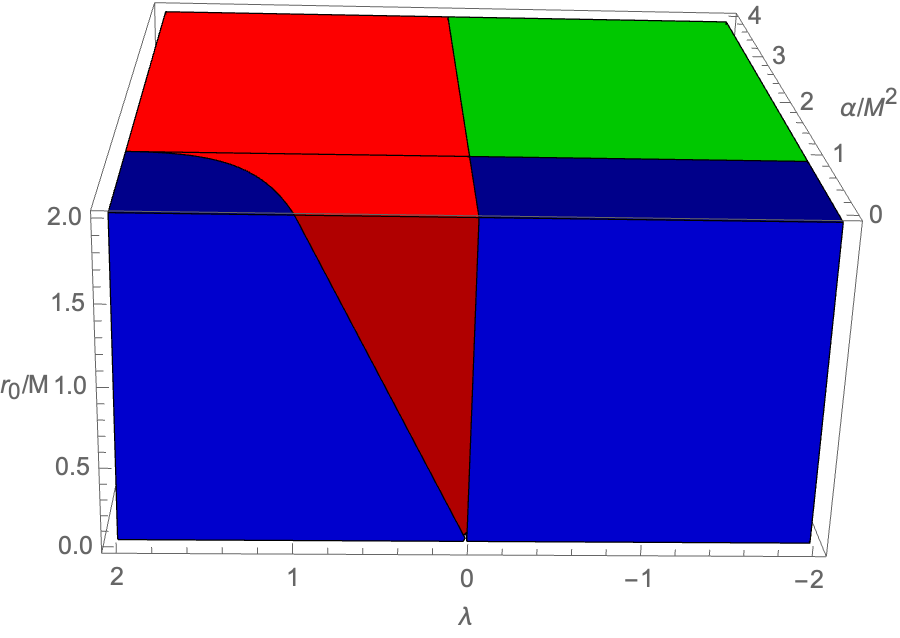} 
\caption{ \it The parameter space of the solutions. 
Choosing the parameters $(\alpha, r_0, \lambda)$ in the red, blue and green regions leads to wormholes, black holes and naked singularities, respectively.}
\label{para}
\end{figure}

If there is neither a black hole horizon nor a wormhole throat, a time-like naked singularity may arise at $r=0$. 
In conclusion, there are three kinds of solutions for compact objects, located in distinct domains in parameter space, satisfying the following conditions,
\begin{itemize}
\item wormholes: a) $\lambda>0$ and $\alpha>M^2$ or b) $\lambda>0, \alpha\leq M^2$ and $r_h<r_t$, then the wormhole throat is located at $r=r_t$;
\item black holes:  a) $\lambda< 0$ and $\alpha\leq M^2$ or b) $\lambda>0, \alpha\leq M^2$, and $r_h\geq r_t$, then the black hole horizon is located at $r=r_h$;
\item naked singularities:
$\lambda< 0$ and $\alpha>M^2$.
\end{itemize}
The parameter space is shown in Fig.~\ref{para}.
Note that for $r_0=0$, we have $g_{tt}=g^{-1}_{rr}$, and the solution reduces to the 4D Gauss-Bonnet black hole of Ref.~\cite{Lu:2020iav,Glavan:2019inb}. 
Moreover, for $\alpha\to 0$ the solution reduces to the Schwarzschild metric.

\subsection{Coordinate transformation}

When we discuss wormhole spacetimes, the coordinates $(t,r,\theta,\phi)$ \eqref{line1}, cover only half of the spacetime since $r \in [r_t,\infty)$.  
Therefore we prefer to choose the radial wormhole coordinate $l$, covering the full spacetime.

The line element then reads in terms of the wormhole coordinate \cite{Bakopoulos:2021liw}
\be\label{whds}
ds^2=-H(l)dt^2+\ft{dl^2}{F(l)}+(l^2+r_t^2)d\Omega^2_2,
\ee
where $l^2=r^2-r_t^2$, 
and the two sides of the wormhole spacetime are given by 
\be l\in (-\infty,0),\qquad l\in (0,+\infty),
\ee
with $l=0$ harboring the throat.
By comparing with the metric functions $h(r)$ and $f(r)$ \eqref{metricf}, one immediately finds the new metric functions $H(l)$ and $F(l)$
\be\label{HF}
H(l)=h(r(l)), \qquad F(l)=\ft{f(r(l))(l^2+r_t^2)}{l^2}.
\ee
For wormholes, the metric functions $H(l)$ and $F(l)$ are both finite and smooth everywhere in the spacetime, $l\in (-\infty,+\infty)$.

In contrast, for black holes, there are two cases. 
For $\lambda>0$, $r_t$ is always a real number and hence the transformation can be employed. 
However, there are no real solutions for $r_t$ when $\lambda\leq0$. 
In this case, we should use the transformation $l^2=r^2-r_h^2$. 
In any case, $l\in (l_h, +\infty)$, where $l_h$ denotes the black hole horizon. 
However, for $r_0\neq 0$, the wormhole coordinate cannot cover the region $r\in (0, r_0)$. 
Hence, we use the original radial coordinate $r$, when we discuss the naked singularity cases.

\section{Geodesics}

For the line element \eqref{whds}, the corresponding first-order geodesic equation for the orbital motion $l(\phi)$ is given by
\be\label{geoeom}
(\ft{dl}{d\phi})^2=(l^2+r_t^2)^2\ft{F(l)}{H(l)}\bigg(\ft{E^2}{L^2}-V(l)\bigg),
\ee
where $E$ and $L$ represent the conserved quantities, energy and angular momentum, respectively, and $V_p(l)$ denotes the effective potential which is given by
\be\label{effVp}
V_p(l)=\ft{H(l)}{l^2+r_t^2}
\ee
for massless particles (photons), and by
\be\label{effVn}
V_n(l)=\ft{H(l)}{l^2+r_t^2}+\ft{H(l)}{L^2}
\ee
for masssive particles.

\subsection{Null geodesics}

We discuss the trajectories of photons first. 
With the help of the impact parameter $b=\ft{L}{E}$, the geodesic equation \eqref{geoeom} becomes
\be\label{geoeomp}
(\ft{dl}{d\phi})^2=(l^2+r_t^2)^2\ft{F(l)}{H(l)}\bigg(\ft{1}{b^2}-V_p(l)\bigg).
\ee
The photon trajectory for a specific impact parameter $b$ has turning points when
\be
\ft{1}{b^2}=V_p(l).
\ee
The light rings, the circular orbits of photons, are the solutions $l=l_{LR}$ of the equations
\be
\ft{\partial V_p}{\partial l}|_{l=l_{LR}}=0, \qquad \ft{\partial^2 V_p}{\partial l^2}|_{l=l_{LR}}<0.
\ee

\subsubsection*{Wormholes:}

In the wormhole case, $H(l)$ and $F(l)$ are positive everywhere. 
There are two types of the effective potential. 
Without loss of generality let us fix 
\be
r_0=1,
\ee
and satisfy $r_t>r_h$. 
The effective potential $V(l)$ has a single peak or a double peak in this case.
We label the single peak wormhole as the type I wormhole and the double peak one as type II wormhole. 
To classify the wormhole solutions, we consider the roots of the following equation \cite{Bakopoulos:2021liw},
\be\label{dVr}
r^3-9M^2 r+8\alpha M=0 .
\ee
If all the roots of \eqref{dVr} are real\footnote{The three real roots $r_1, r_2, r_3$ correspond to the locations of the extrema.}, we have a type II wormhole, and otherwise a type I wormhole.

\begin{figure}[h]
\centering
\includegraphics[width=0.47\textwidth]{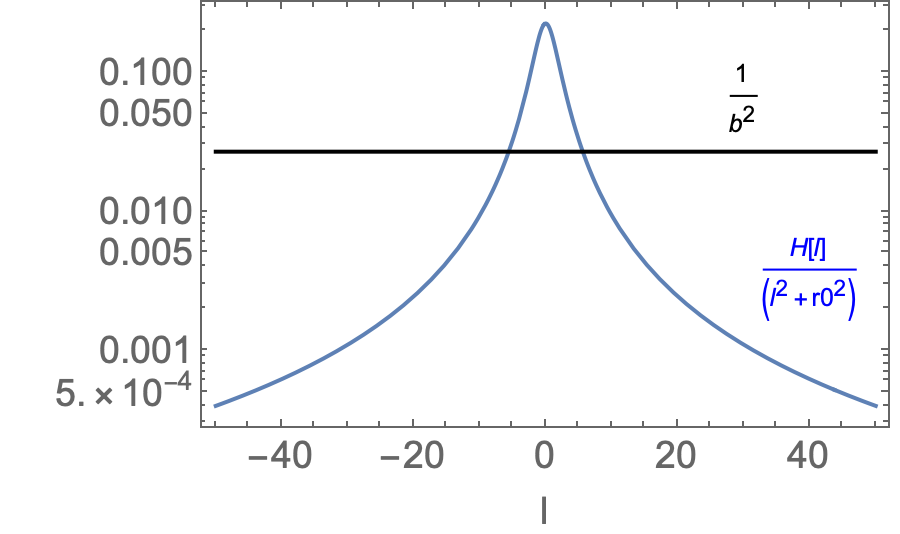} 
\includegraphics[width=0.45\textwidth]{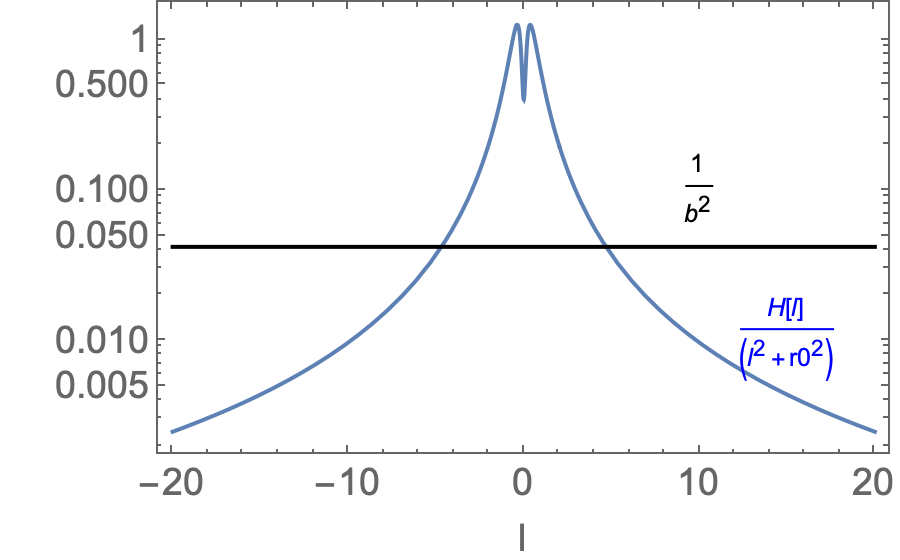} 
\caption{ \it The effective potential for null geodesics $V_p(l)$ in the wormhole case. 
The left one is for the type I wormhole featuring a single peak. 
The right one is for the type II wormhole and has a double peak.}
\label{plotVWH}
\end{figure}

\begin{figure}[h]
\centering
\includegraphics[width=0.4\textwidth]{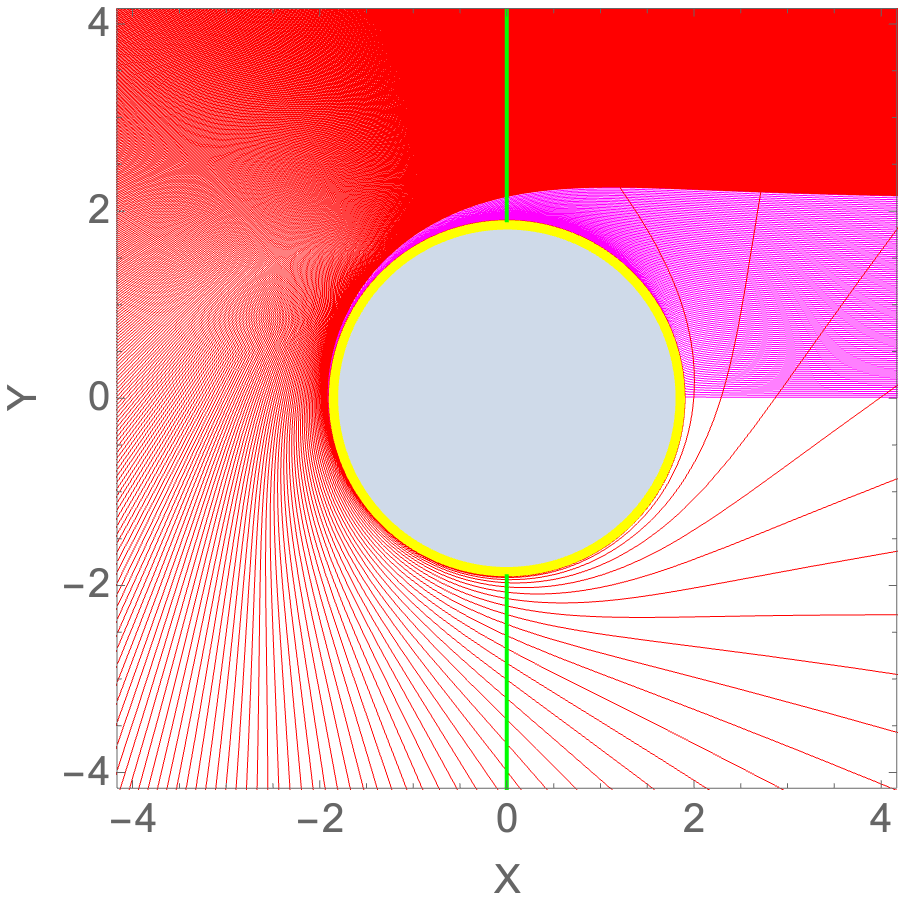} 
\includegraphics[width=0.42\textwidth]{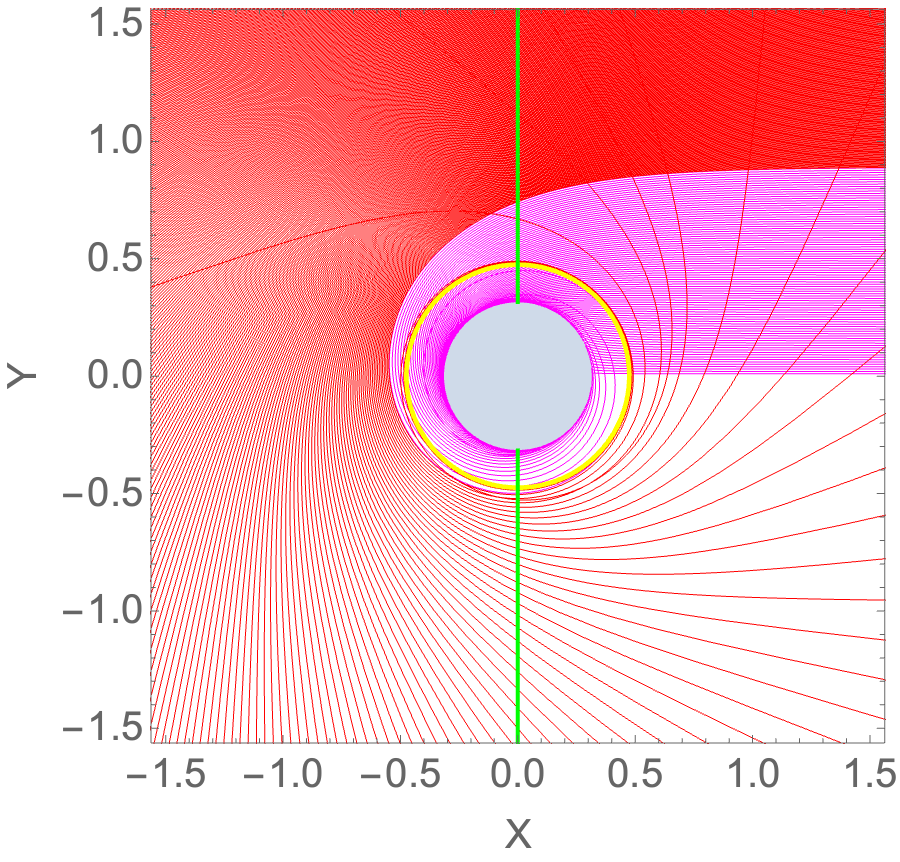} 
\caption{ \it Wormhole geodesics: 
The corresponding null geodesics for the effective potentials in Fig.~\ref{plotVWH}. 
The left one is for the type I wormhole, while the right one is for the type II wormhole. 
The red lines denote the light rays which do not enter the wormhole throat.
The purple lines denote the light rays that cross the wormhole throat. 
The yellow circle represents the light ring. 
The green lines denote the accretion disks.}
\label{plotgeo}
\end{figure}

For the type I wormhole with a single peak 
the single light ring coincides with the wormhole throat at $l=0$. 
In the numerically calculated example, we chose the parameters $\alpha=0.02, M=0.181, \lambda=0.1$. 
The corresponding effective potential is shown in the left plot of Fig.~\ref{plotVWH}.

For the type II wormhole with a double peak there are two unstable light rings, located on both sides of the wormhole throat. 
The height of the two peaks is the same, because the wormhole is symmetric. 
The parameters for the numerically calculated example are given by $\alpha=0.02, M=0.181, \lambda=0.8$ in this case.
The light rings are then located at $l_{LR}=\pm 0.3674$. 
We show the corresponding effective potential in the right plot of Fig.~\ref{plotVWH}.

It is straightforward to solve the geodesic equation \eqref{geoeom} numerically. 
The results are shown in Fig.~\ref{plotgeo} using the celestial coordinates $X=r \cos\phi, Y=r \sin\phi$. 
The critical impact parameter is $b_c=2.1147$ for the type I wormhole and $b_c=0.8918$ for the type II wormhole. 
The light rays with $b>b_c$ are  bent by the wormhole.
They are depicted in red. 
Otherwise, the light rays cross the wormhole throat to the other universe. These are depicted in purple.

\subsubsection*{Black holes:}

\begin{figure}[h]
\centering
\includegraphics[width=0.47\textwidth]{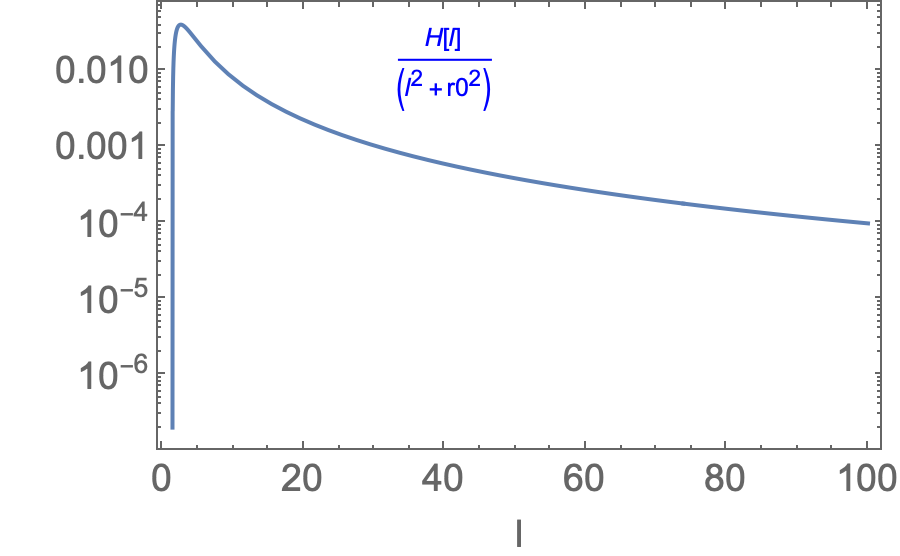} 
\includegraphics[width=0.45\textwidth]{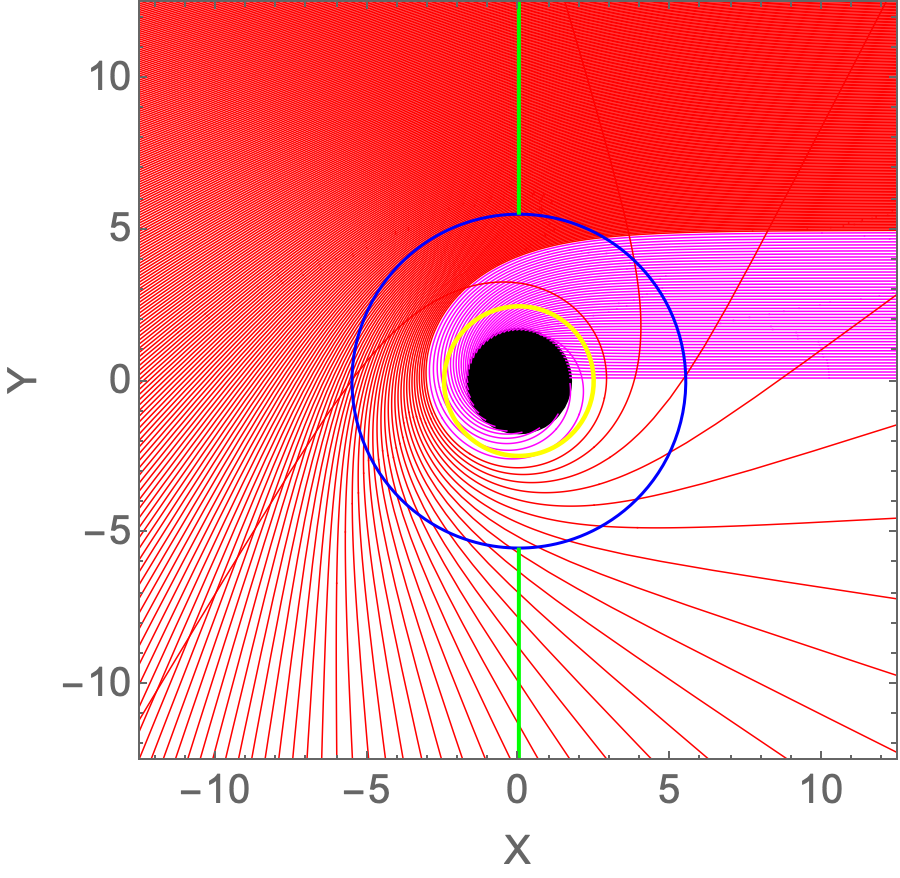} 
\caption{ \it Left plot: the effective potential for null geodesics $V_p(l)$ in the black hole case. 
Right plot: the black hole null geodesics  of the effective potential in Fig.~\ref{plotVBH} (left). 
The red lines denote the light rays which do not fall into the black hole.
The purple lines denote the light rays which enter
the horizon. The yellow circle represents the light ring. 
The blue circle represents the ISCO. 
The green lines denote the accretion disks.}
\label{plotVBH}
\end{figure}

In the black hole case, there is a horizon instead of a throat. 
We recall that the metric \eqref{whds} then reduces to the 4D Gauss-Bonnet black hole metric if we set $r_0=0$ \cite{Glavan:2019inb,Lu:2020iav}. 
To present a numerical example, we chose the parameters $r_0=1, M=1, \alpha=0.523, \lambda=0.75$. 
The resulting black hole has two horizons and a light ring outside the outer horizon. 
The corresponding effective potential $V_p(l)$ is shown in the left plot of Fig.~\ref{plotVBH}. 
By inspecting the effective potential, one can infer that its properties are similar to those of the Schwarzschild or Reissner-Nordstr\"om black holes, because of the similar shapes of their effective potentials. 
We show the numerical results for the null geodesics in the right plot of Fig.~\ref{plotVBH}, which confirms the above inference. 
Here the outer horizon is located at 
\be
l_h=1.3632,
\ee
and the light ring at 
\be
l_{LR}=2.5434,
\ee
with the corresponding critical impact parameter $b_c=4.9708$.

\subsubsection*{Naked singularities:}

In the case of naked singularities there are no more horizons. 
As discussed before, we need to use the original radial coordinate $r$ in that case.
The effective potential is then given by
\be
V_p(r)=\ft{h(r)}{r^2}.
\ee
There are two possible shapes for the effective potential $V_p(r)$ in the case of naked singularities. 

\begin{figure}[h]
\centering
\includegraphics[width=0.47\textwidth]{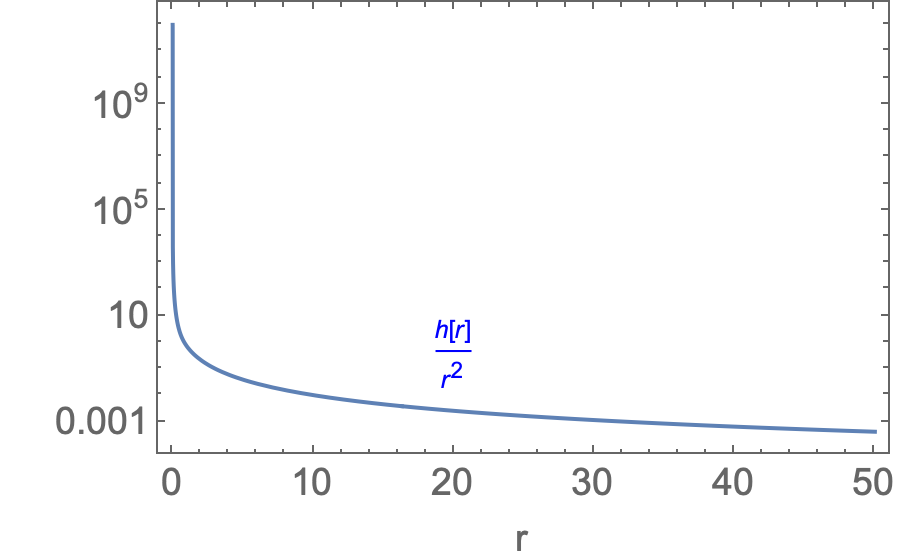} 
\includegraphics[width=0.45\textwidth]{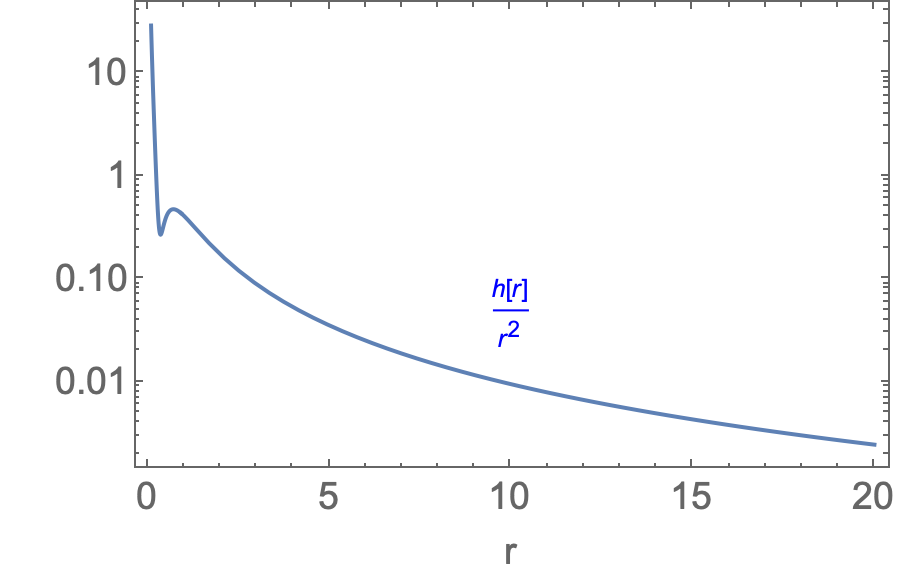} 
\caption{ \it The effective potential for null geodesics $V_p(r)$ in the naked singularity case. 
The left one has no light ring. 
The right one features an unstable light ring.}
\label{plotVNS}
\end{figure}

We refer to a type I naked singularity when the metric does not feature a light ring. 
The effective potential then decreases monotonically from infinity to zero, when $r$ runs from zero to infinity. 
To illustrate this with an example, we take the parameters $\alpha=0.523, M=0.181, r_0=0.1, \lambda=-0.3$ and show it in the left plot of Fig.~\ref{plotVNS}.

In the case of a type II naked singularity, the effective potential does have a light ring.
The effective potential is then no longer monotonically decreasing.
Instead it has a local maximum and minimum. 
The local maximum relates to the unstable light ring. 
Here we choose the parameters $\alpha=0.11, M=0.316, r_0=0.1, \lambda=-0.3$ as an example and exhibit it in the right plot of Fig.~\ref{plotVNS}. 
The unstable light ring is located at $r=0.7145$.

\begin{figure}[h]
\centering
\includegraphics[width=0.42\textwidth]{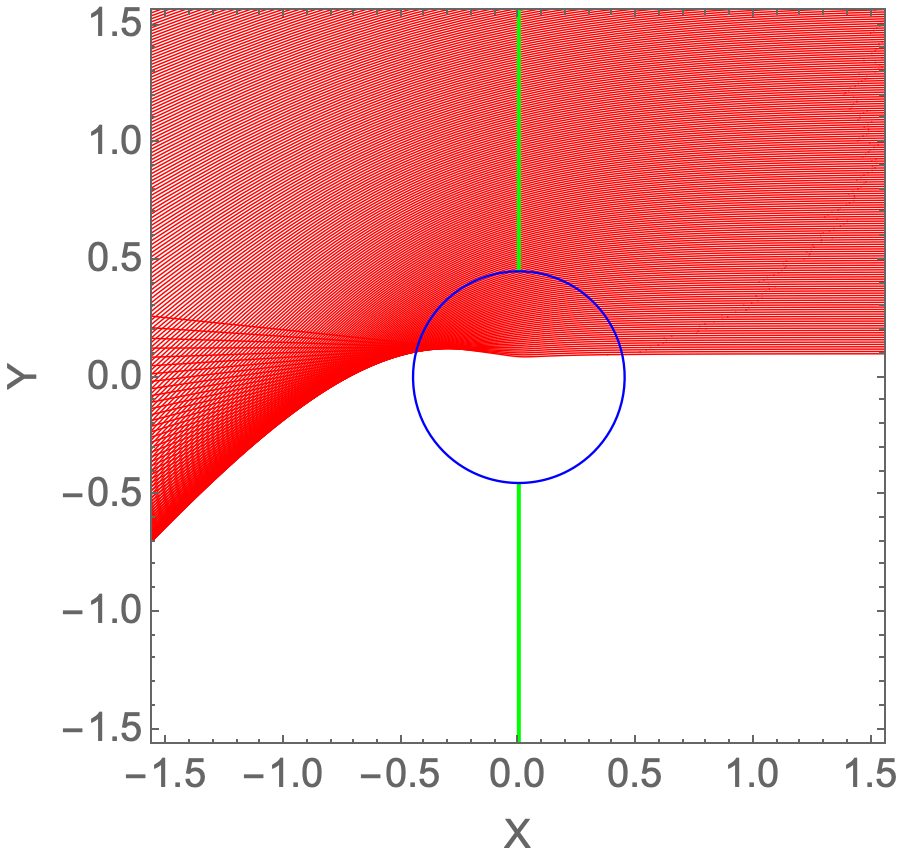} 
\includegraphics[width=0.4\textwidth]{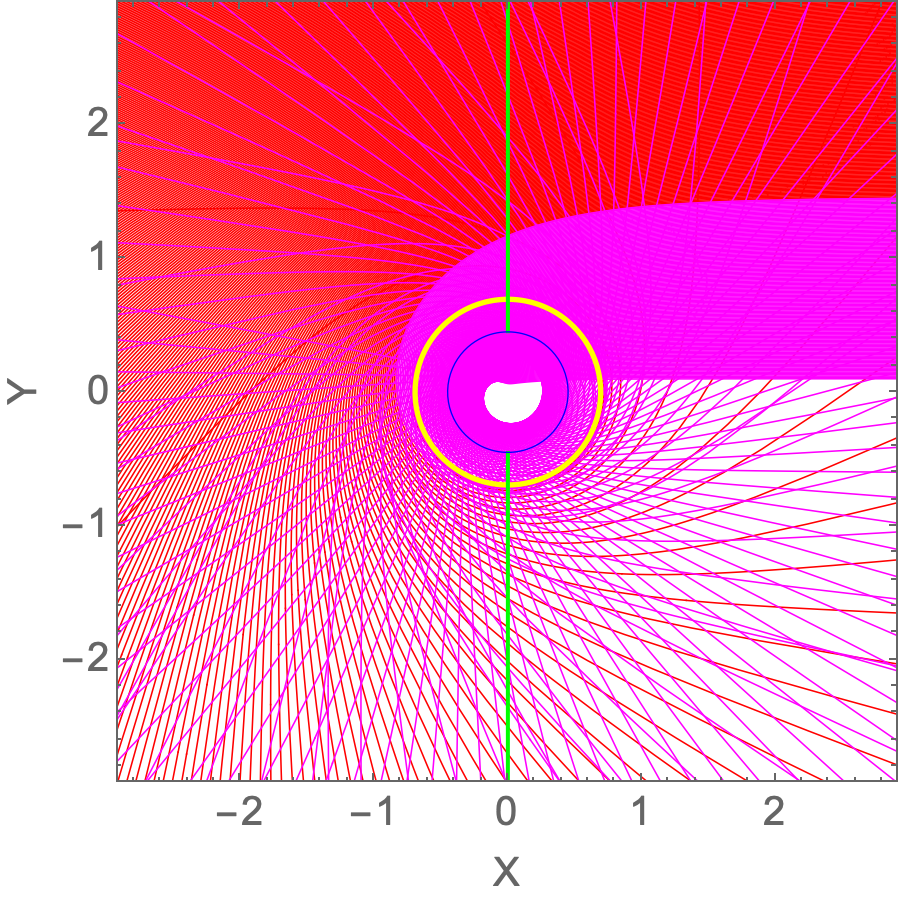} 
\caption{ \it Naked singularity geodesics: 
The corresponding null geodesics for the effective potentials of Fig.~\ref{plotVNS}. 
The red lines denote the light rays which do not cross the unstable light ring, if present. 
The purple lines denote the light rays that do cross this light ring. 
The yellow circle represents the light ring. 
The blue circles indicate the static spheres (see subsection 3.2).
The green lines denote the accretion disks.}
\label{plotgeoNS}
\end{figure}

The corresponding null geodesics for these two types of naked singularities are shown in Fig.~\ref{plotgeoNS}. 
The left one is for the naked singularity without a light ring.
Here the light rays are not that much bent by the singularity. In contrast, for the naked singularity with an unstable light ring, shown in the right plot of Fig.~\ref{plotgeoNS}, the light rays can even move on circular orbits.

\subsection{Bound orbits for massive particles}

Next we discuss the effective potential $V_n$ for massive particles. 
When considering wormholes, black holes and naked singularities,
we are only interested in asymptotically flat spacetimes.
Thus, at large $l$ (or $r$), we find for the effective potential
\be
V_n(l) \sim \ft{1}{L^2}+\cdots.
\ee

\subsubsection*{Wormholes:} 

For wormholes we find close to the throat ($l = 0$),
\be
V_n(0) \sim H(0)\bigg(\ft{1}{L^2}+\ft{1}{r_t^2}\bigg)+\cdots.
\ee
Furthermore, we note that $\ft{\partial V}{\partial l}|_{l=0} = 0 $, since the wormhole is symmetric. 
Thus there is a maximum or a minimum at $l=0$, depending on the sign of the second derivative of the effective potential
\be\label{V0}
\ft{\partial^2 V}{\partial l^2}|_{l=0}=H''(0)\bigg(\ft{1}{L^2}+\ft{1}{r_t^2}\bigg)-\ft{2H(0)}{r_t^4}.
\ee
In addition, there is a critical angular momentum $L_c$, such that $\ft{\partial^2 V}{\partial l^2}|_{l=0}=0$, when $L=L_c$.

\begin{figure}[t]
\centering
\includegraphics[width=0.3\textwidth]{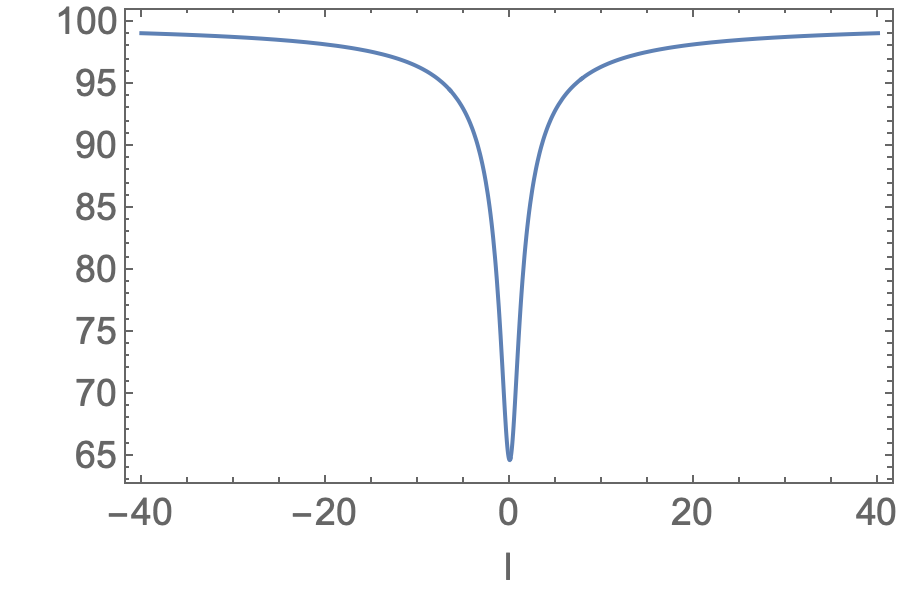} 
\includegraphics[width=0.3\textwidth]{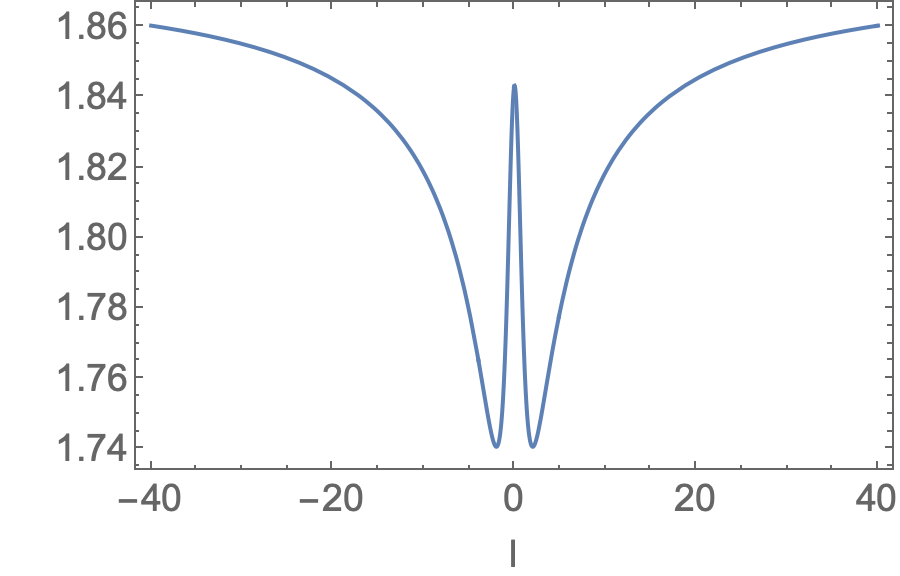} 
\includegraphics[width=0.3\textwidth]{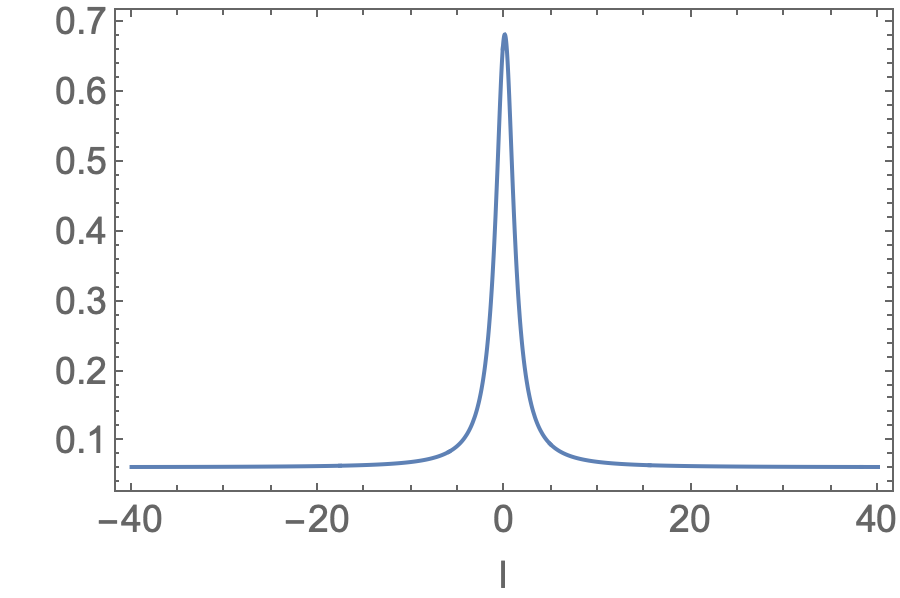} 
\includegraphics[width=0.3\textwidth]{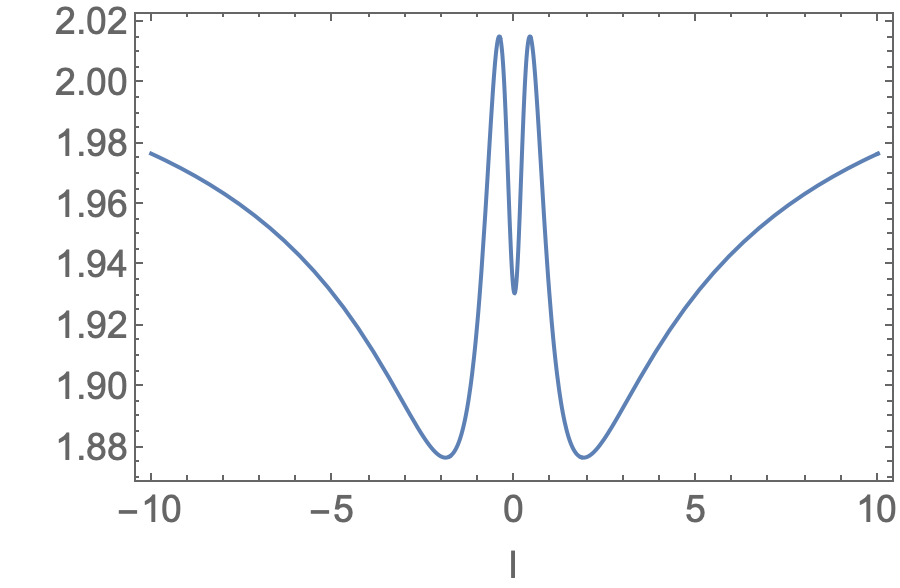} 
\includegraphics[width=0.3\textwidth]{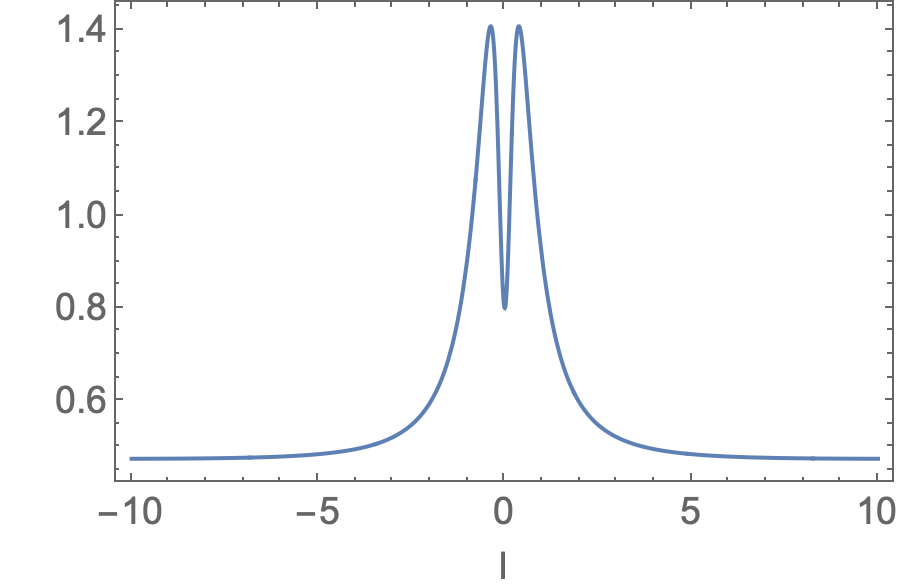} 
\caption{ \it The effective potential $V_n(l)$ for time-like geodesics for the wormhole cases. 
We chose $\alpha=0.02, M=0.181$ and $(L, \lambda)=(0.1,0.2), (0.73,0.2), (4,0.2), (0.7,0.42), (1.44,0.7)$ from left to right and top to down.}
\label{iscowh}
\end{figure}

We show the effective potential $V_n(l)$ for the wormhole cases in Fig.~\ref{iscowh}. 
There are five possible shapes of the effective potential, as illustrated in the figure. 
We note, that when increasing the angular momentum $L$ from less than $L_c$ to greater than $L_c$, the shape of the effective potential changes from the upper left figure to the upper middle one. 

There is an innermost bound orbit at the wormhole throat $l=0$ for massive particles with a suitable angular momentum $L$. 
Following Refs.~\cite{Grandclement:2014msa,Teodoro:2020kok}, we call it ``Innermost Circular Orbit (ICO)". 
Since the ICO is an innermost bound orbit for massive particles, we naively consider the accretion disk extending from the ICO in these wormhole cases.

\subsubsection*{Black holes:} 

\begin{figure}[h]
\centering
\includegraphics[width=0.3\textwidth]{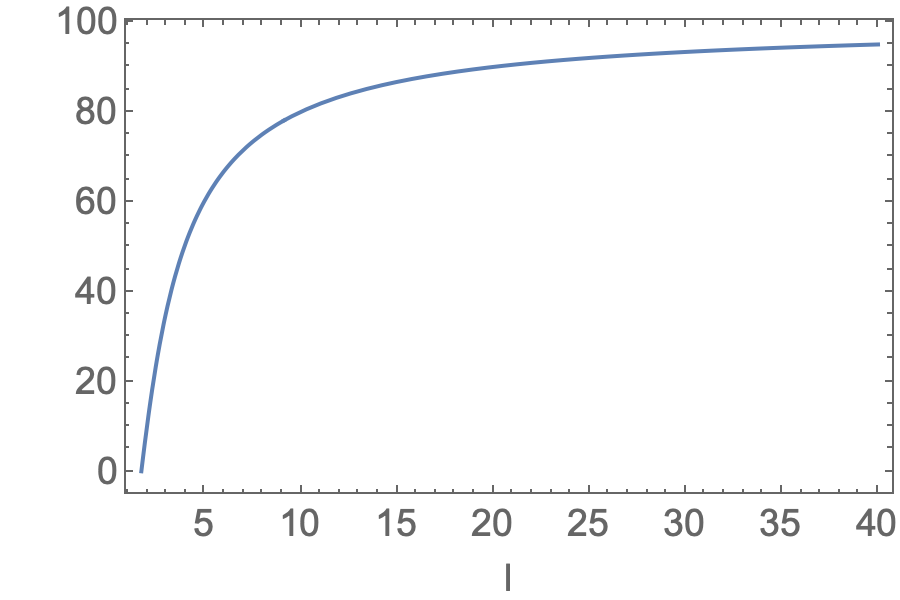} 
\includegraphics[width=0.3
\textwidth]{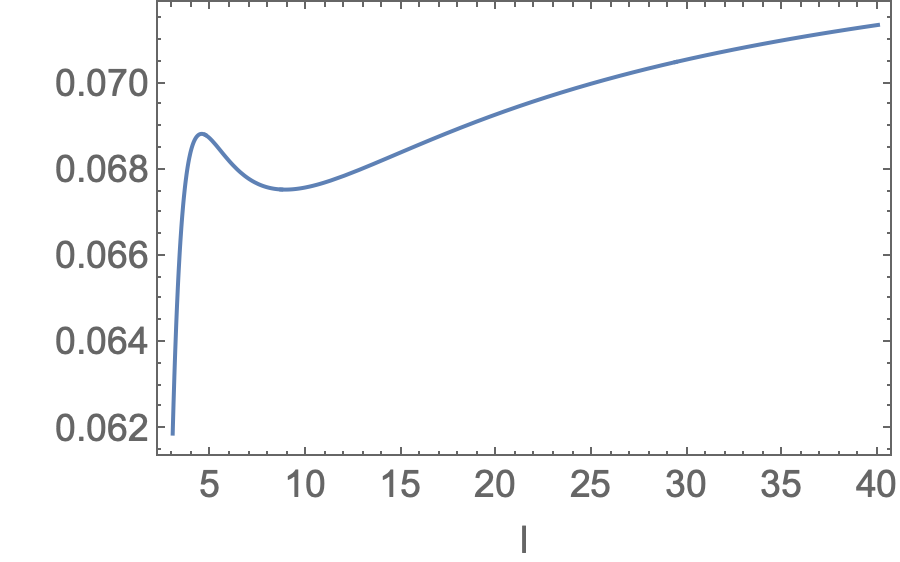} 
\includegraphics[width=0.3
\textwidth]{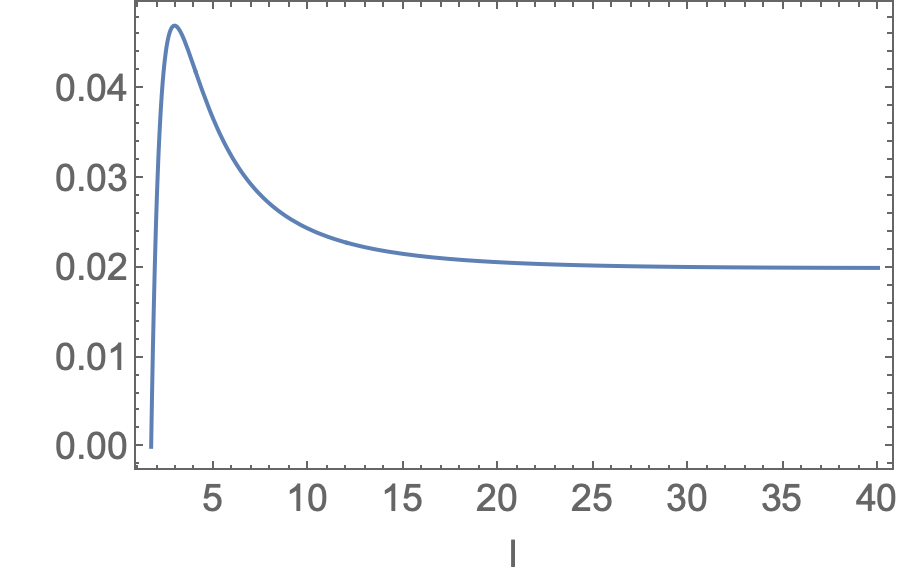} 
\caption{ \it The effective potential $V_n(l)$ for time-like geodesics for the black hole cases. 
We chose $(\alpha, M, L)=(0.523, 1, 0.1), (0.01, 1, 3.66), (0.523, 1, 7)$ from left to right.}
\label{iscobh}
\end{figure}

For the black hole cases, the shapes of the effective potential $V_n(l)$ share similar properties with the Schwarzschild black holes or the Reissner-Nordstr\"om black holes, qualitatively. There is an innermost stable circular orbit (ISCO) with a certain angular momentum $L$ for massive particles. 
We show the three possible shapes for the effective potential $V_n(l)$ in Fig.~\ref{iscobh}. 
The location of the ISCO can be expressed as\footnote{It may, e.g., be obtained by performing a coordinate transformation in Eq.~(4.6) of Ref.~\cite{Gao:2023mjb}.}
\be
l_{ISCO}=\sqrt{\bigg(\ft{3l_{ISCO}^2\sqrt{l_{ISCO}^2+r_t^2}HH'}{H(l_{ISCO}(l_{ISCO}^2+r_t^2)H''-H')-2l_{ISCO}(l_{ISCO}^2+r_t^2)H'^2}\bigg)^2-r_t^2}.
\ee
For the parameter choice $r_0=0, M=1, \alpha=0.523$ the location of the ISCO is $l_{ISCO}=5.6467$.
Like many references (see e.g.~\cite{Gralla:2019xty}), we assume in the black hole cases that the accretion disk extends from the ISCO.

\subsubsection*{Naked singularities:} 

\begin{figure}[h]
\centering
\includegraphics[width=0.3\textwidth]{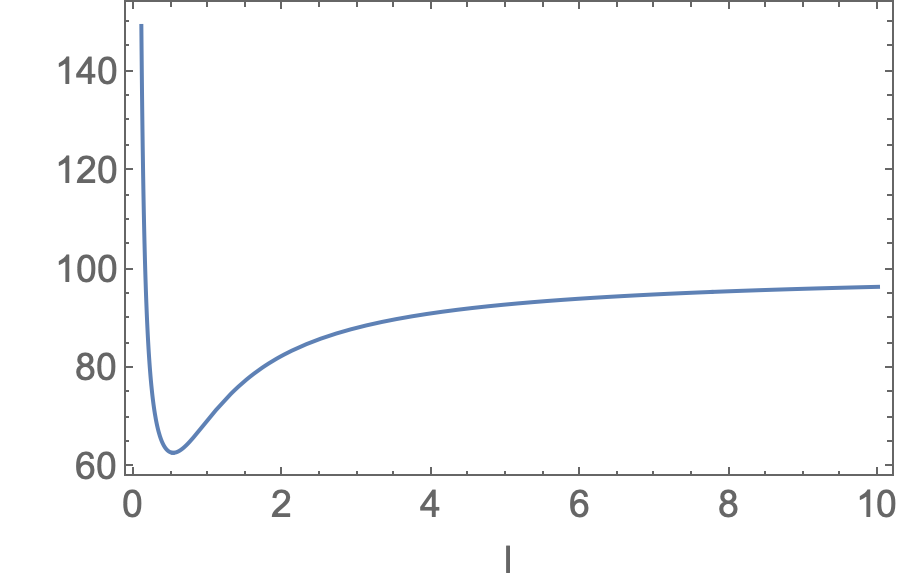} 
\includegraphics[width=0.3
\textwidth]{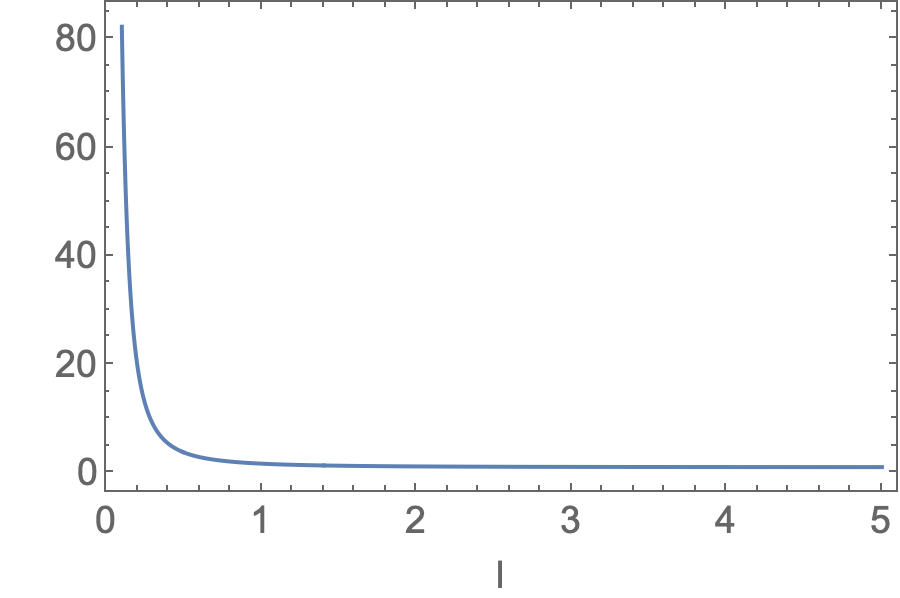} 
\caption{ \it The effective potential $V_n(l)$ for time-like geodesics for the naked singularity cases. 
We chose $(\alpha, M, L)=(0.523, 0.181, 0.1), (0.523, 0.1, 1)$ from left to right.}
\label{iscons}
\end{figure}

For the naked singularity cases, the effective potential $V_n(r)=\ft{h(r)}{r^2}+\ft{h(r)}{L^2}$ diverges, when $r\to 0$. 
The two shapes of the effective potential are shown in Fig.~\ref{iscons}. 

For a given $(\alpha, M)$\footnote{Recall that $\alpha>M^2$ for a naked singularity.
}, there is a special sphere at
\be
r_{ss}=M^{1/3}\alpha^{1/3},
\ee
such that massive particles with vanishing angular momentum ($L=0$) and energy
\be
E=\sqrt{1-\ft{M^{2/3}}{\alpha^{1/3}}}
\ee
can stay at rest. 
Following Ref.~\cite{Wei:2023bgp}, we call this sphere the ``static sphere" of the naked singularity. 
We assume that the accretion disk extends from the static sphere in the naked singularity cases.
Taking the same parameter values as in the previous subsection, the static sphere for the type I naked singularity is located at $r_{ss}=0.4558$ and for the type II singularity at $r_{ss}=0.3264$.

\section{Shadow images with an accretion disk}

To obtain the images of the wormholes, black holes and the naked singularities, we assume that these compact objects are illuminated by an optically and geometrically thin disk. 
This light source, that is extending from the ICO (for wormholes), the ISCO (for black holes) or the static sphere (for naked singularities), is marked by a respective green line in the above figures of the null geodesics. 

The distribution of the emitted light intensity is defined
by the function $I_{em}=I_{em}(l)$.
According to the standard techniques employing the so-called ``ray-tracing method", the observed intensity $I_{obs}(b)$ is obtained from the emitted intensity via
\be\label{inten}
I_{obs}(b)= 
\sum_n (g_{tt}(r))^2\, I_{em}|_{r=r_n(b)} ,
\ee
where $r_n(b)$ is the transfer function that indicates the $n^{th}$ intersection with the accretion disk.

\subsection*{Wormholes:}

For the wormhole cases, the accretion disk extends from the wormhole throat. Hence, we choose the following emitting intensity function,
\begin{eqnarray}
&&I_{em}=
\begin{cases}
\frac{I_0}{(-l-5)^2}, & l\leq 0\\ \nn
\frac{I_0}{(l-5)^2}, &  l>0,
\end{cases}
\label{intensityWH}
\end{eqnarray}
where $I_0$ is a constant that adjusts the strength of the emitted intensity.  
\begin{figure}[h]
\centering
\includegraphics[width=0.4\textwidth]{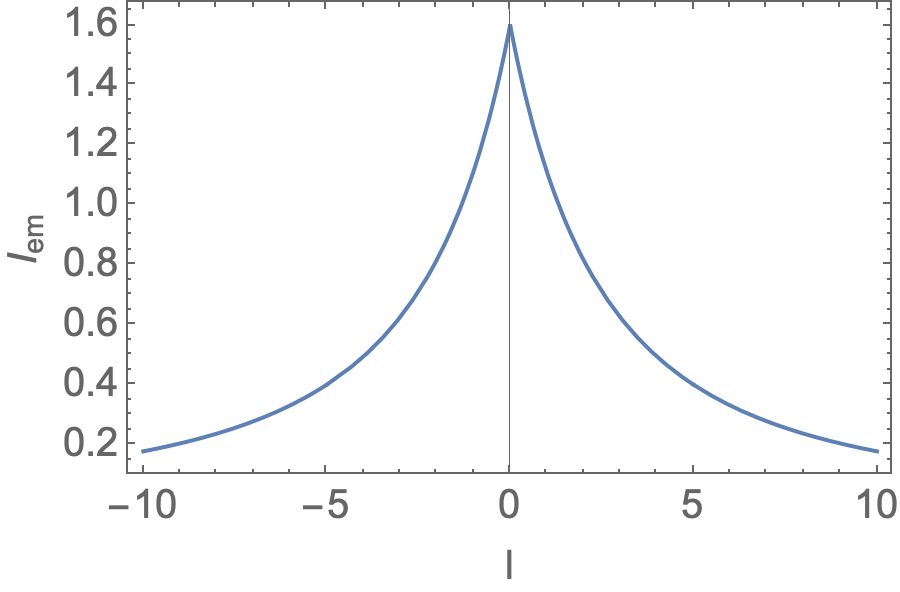} 
\includegraphics[width=0.42\textwidth]{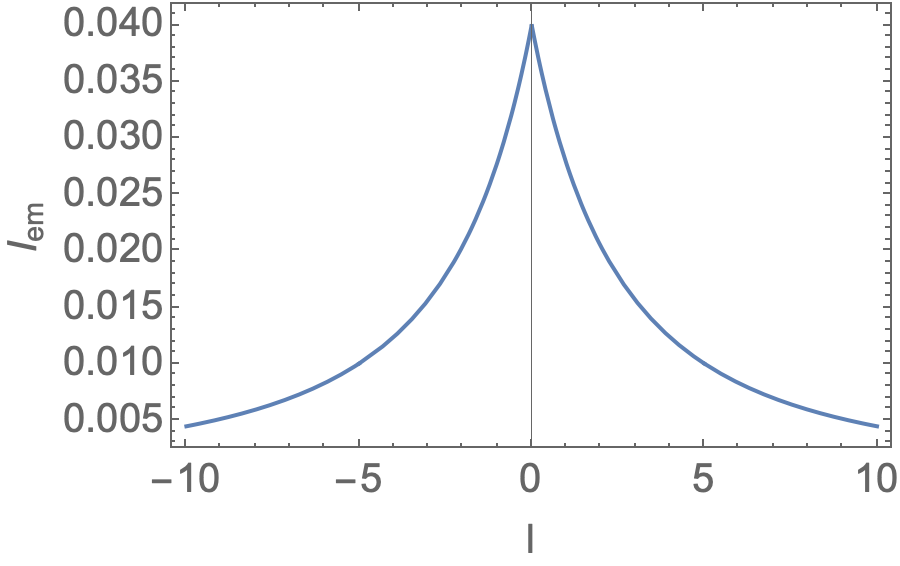} 
\caption{ \it The emitted intensity as a function of $l$. 
The left plot has $I_0=40$ for a type I wormhole, and the right plot $I_0=1$ for a type II wormhole.}
\label{Iemwh}
\end{figure}

\begin{figure}[h]
\centering
\includegraphics[width=0.42\textwidth]{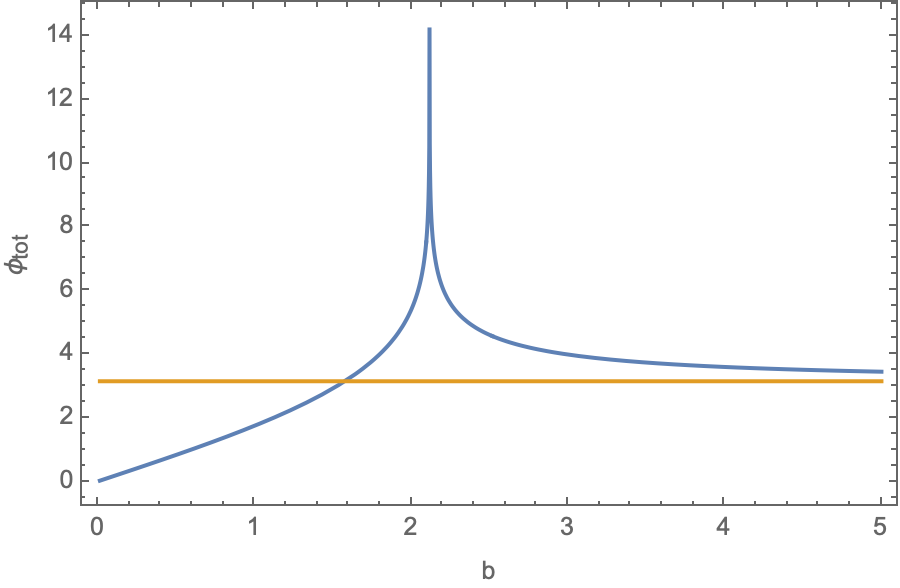} 
\includegraphics[width=0.42\textwidth]{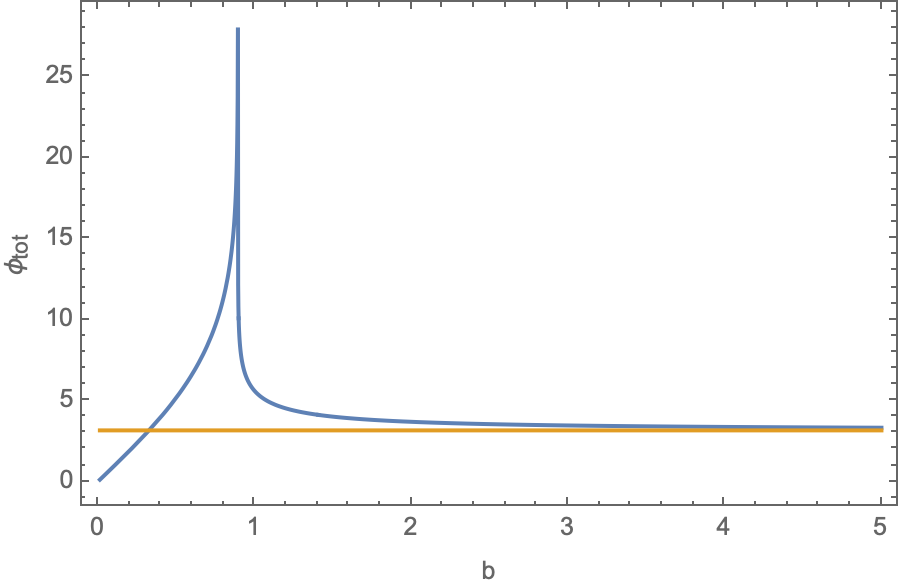} 
\caption{ \it The total deflection angle as a function of the impact parameter $b$. 
The horizontal line in orange indicates $\phi=\pi$. 
The left plot is for a type I wormhole and the right one for a type II wormhole.}
\label{phitotwh}
\end{figure}

\begin{figure}[h]
\centering
\includegraphics[width=0.4\textwidth]{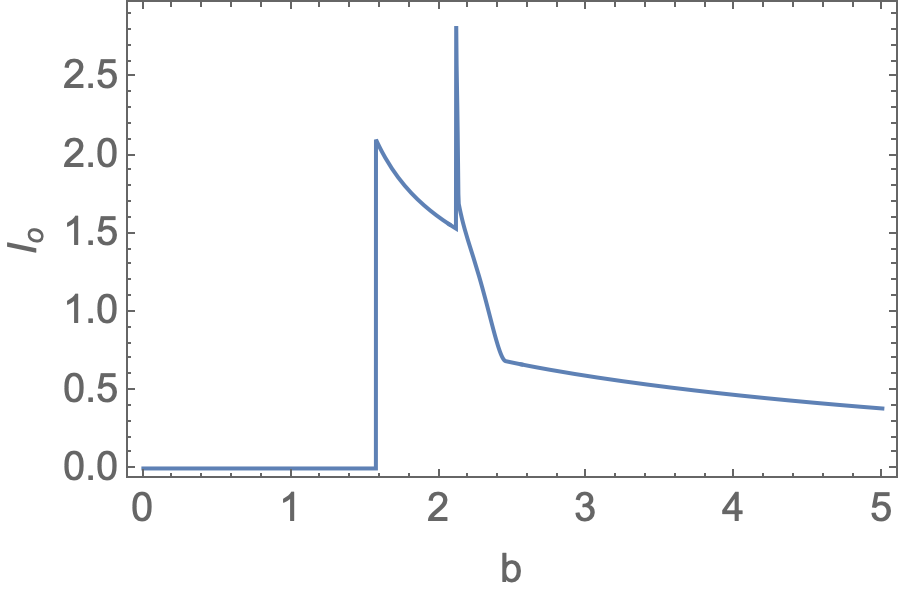} 
\includegraphics[width=0.42\textwidth]{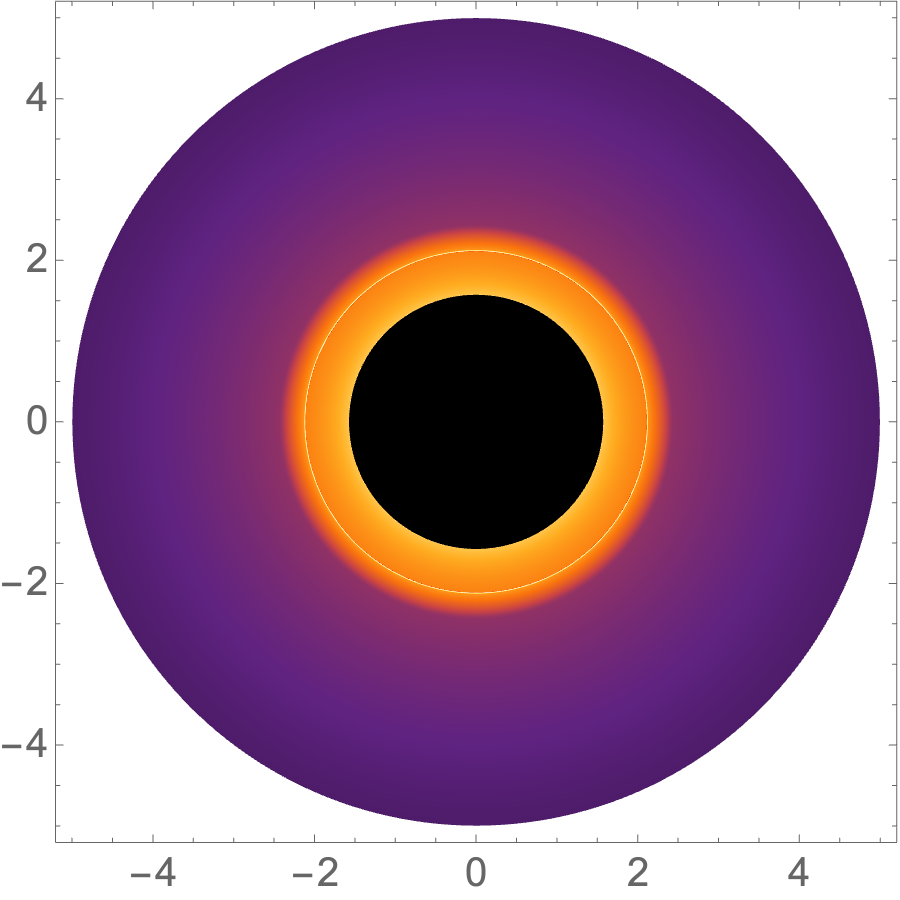} 
\caption{ \it The observed intensity and the corresponding optical image of a type I wormhole.}
\label{imagewh1}
\end{figure}

\begin{figure}[h]
\centering
\includegraphics[width=0.4\textwidth]{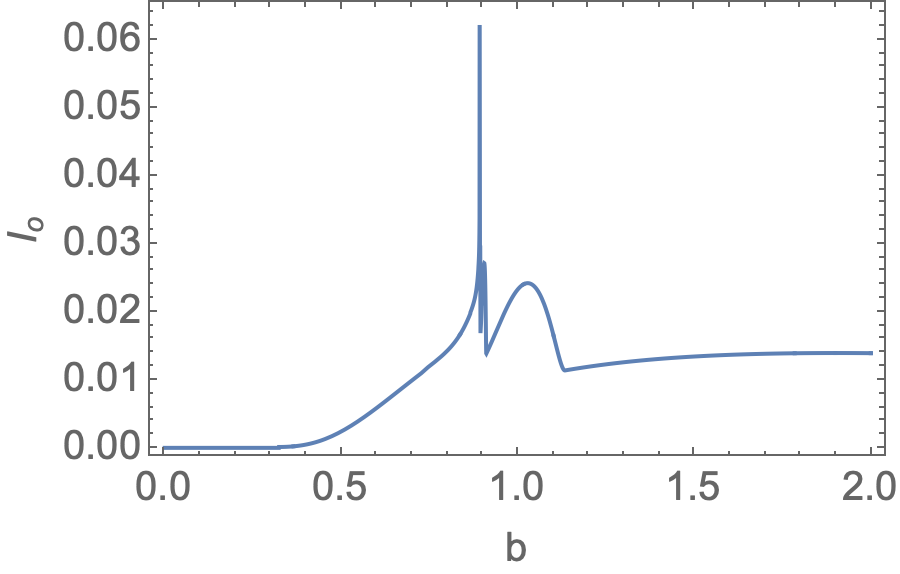} 
\includegraphics[width=0.42\textwidth]{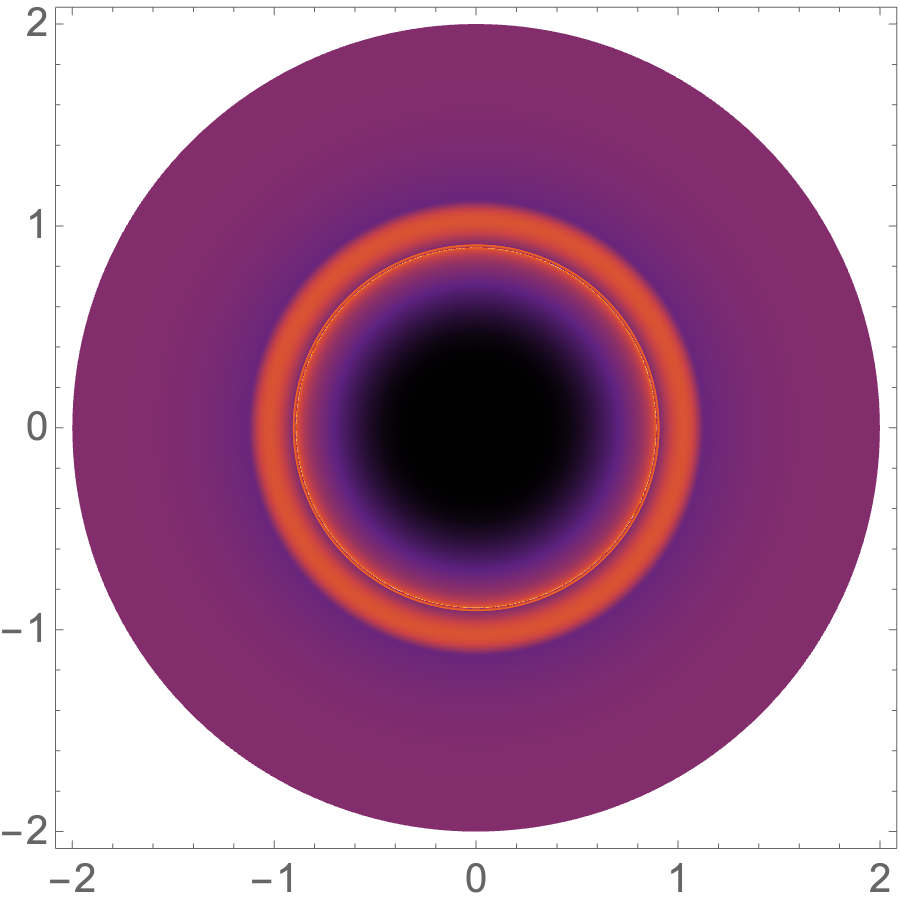} 
\caption{ \it The observed intensity and the corresponding optical image of a type II wormhole.}
\label{imagewh2}
\end{figure}

For the type I wormhole, whose light ring is located at the wormhole throat, we set $I_0=40$ as shown in the left plot of Fig.~\ref{Iemwh}. 
The deflection angle is shown as a function of the impact parameter $b$ in the left plot of Fig.~\ref{phitotwh}. 
Since the light ring is located at the wormhole throat, the deflection angle tends to infinity there. 
The observed intensity and the corresponding optical image are shown in Fig.~\ref{imagewh1}. 
The shadow is cast by the light rays, that never hit the accretion disk. Since the accretion function \eqref{intensityWH} decreases faster with increasing distance from the wormhole than the metric function $g_{tt}$ increases, the observed intensity decreases. 
The light ring leads to a bright ring outside the shadow.

For the type II wormhole, there are two unstable light rings, located symmetrically with respect to the throat. 
We set $I_0=1$ in this case and depict the emitted intensity in the right plot of Fig.~\ref{Iemwh}. 
The deflection angle is shown in the right plot of Fig.~\ref{phitotwh}. 
It is similar to the one of the type I wormhole, tending to infinity at the light ring. 
We present the observed intensity and the corresponding optical image in Fig.~\ref{imagewh2}. 
Since the accretion disk extends inside the unstable light rings, this leads to an increase of the observed intensity at first. Again, a bright ring arises outside the shadow.

One of the most interesting features observed when comparing the images of type I and type II wormholes, is that the boundary of their shadow is different. The shadow of a type I wormhole has a clear boundary, but this does not hold for the shadow of a type II wormhole.

\subsection*{Black holes:}

\begin{figure}[h]
\centering
\includegraphics[width=0.42\textwidth]{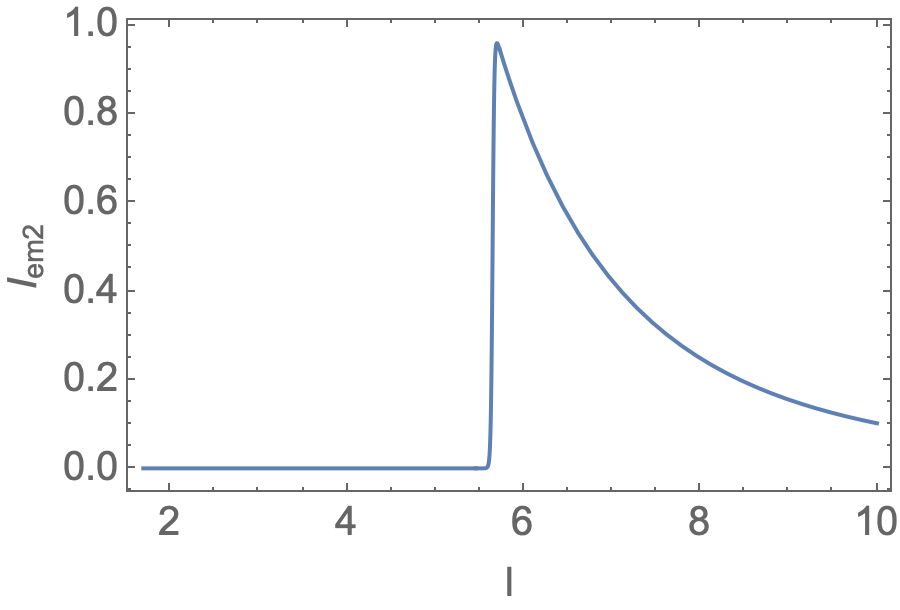} 
\includegraphics[width=0.42\textwidth]{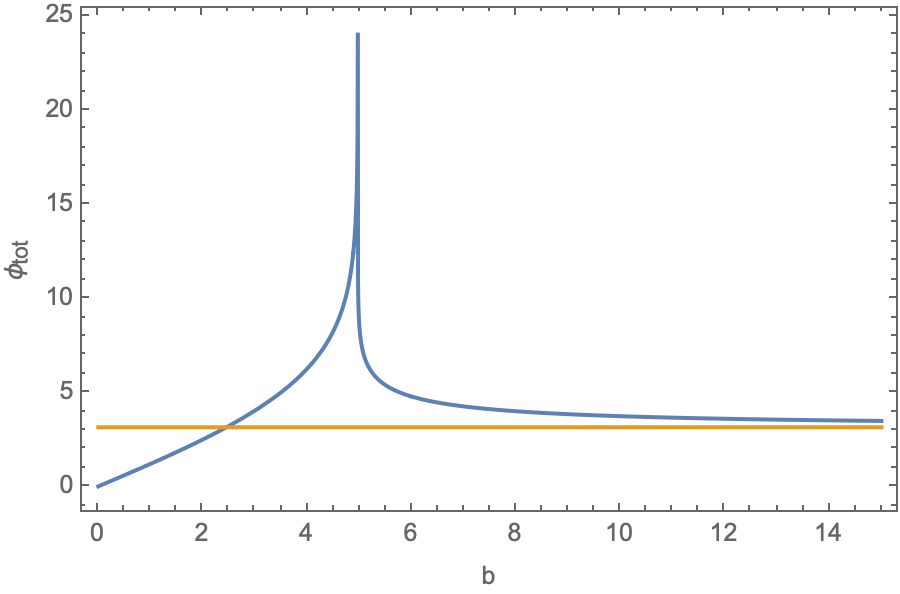} 
\caption{ \it Left: the emitted intensity as a function of $l$ for black hole case. Right: the total deflection angle as a function of $b$.}
\label{Iembh}
\end{figure}

For the black hole case we adopt the following model for the emitting intensity
\be
I_{em}=(\ft{l_{ISCO}}{l})^4\ft{1+\tanh\big(50(l-l_{ISCO})\big)}{2},
\ee
where $l_{ISCO}=5.6467$ is the location of the ISCO. 
As seen in the left plot of Fig.~\ref{Iembh}, the function $I_{em}$ decreases from the ISCO. 
The deflection angle is shown as a function of the impact parameter $b$ in the right plot of Fig.~\ref{Iembh}. 

\begin{figure}[h]
\centering
\includegraphics[width=0.4\textwidth]{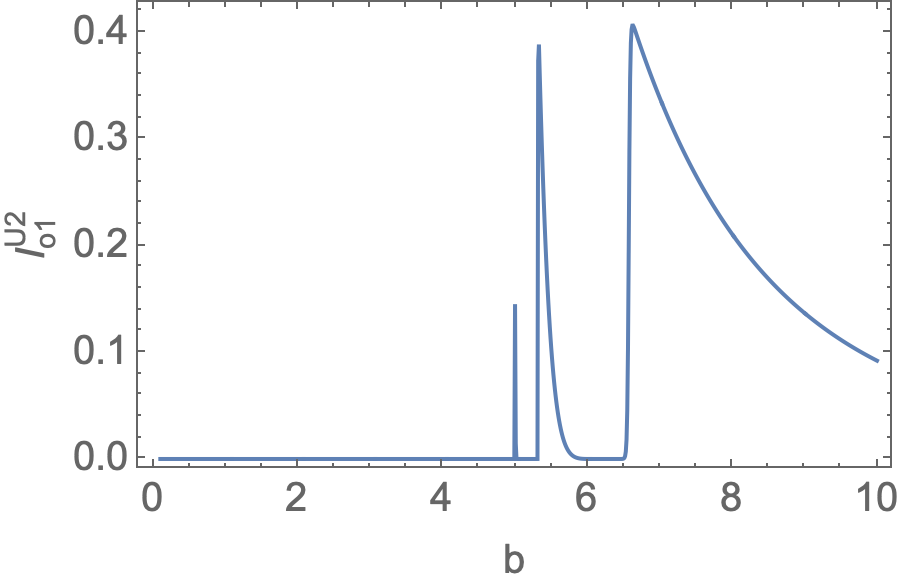} 
\includegraphics[width=0.42\textwidth]{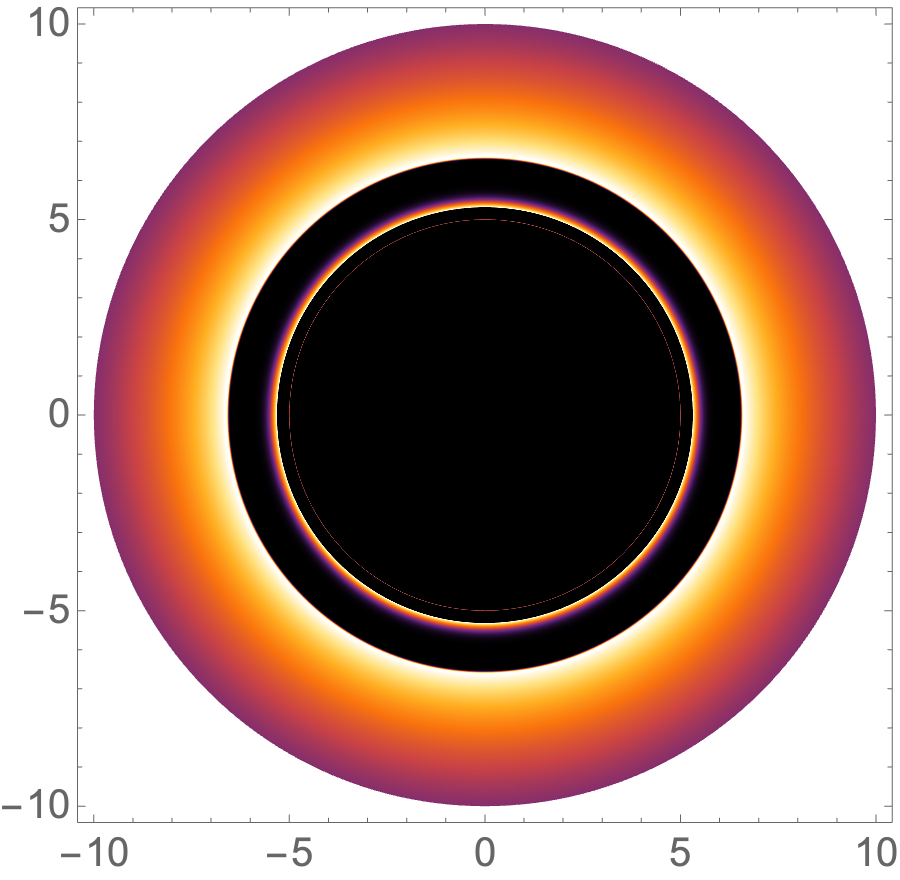} 
\caption{ \it The observed intensity and the corresponding optical image of the black hole.}
\label{imagebh}
\end{figure}

The observed intensity $I_{obs}$ and the corresponding optical image are shown in Fig.~\ref{imagebh}. 
Here we see a bright ring inside the shadow. 
In fact, as seen in the observed intensity, there should be many bright rings inside the shadow. 
The reason is that the closer the light rays are located to the light ring, the more often they circle the black hole.
In fact, each number of times a light ray circles the black hole, gives rise to a bright ring. At the same time, the gaps between the corresponding impact parameters are getting smaller and smaller, so that we can not observe them separately. 
These results are similar to those for typical black holes, like Schwarzschild black holes \cite{Gralla:2019xty}. Because the effective potential $V_p$ only  involves the metric function $g_{tt}$, one can infer that the black hole images with $r_0=0$, i.e. for the black holes in 4-D Gauss-Bonnet gravity\cite{Lu:2020iav,Glavan:2019inb}, are similar to ours.


\subsection*{Naked singularities:}

\begin{figure}[t]
\centering
\includegraphics[width=0.32\textwidth]{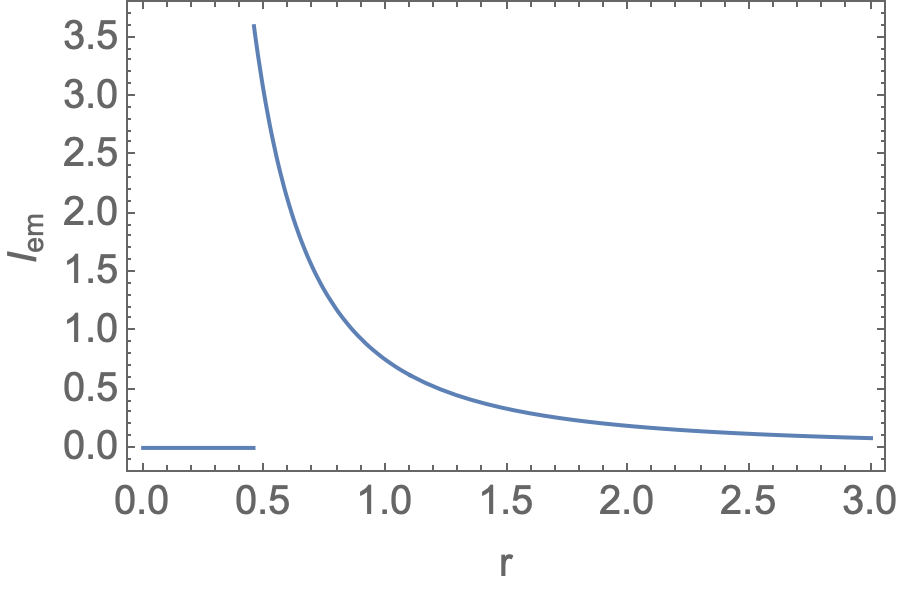} 
\includegraphics[width=0.32\textwidth]{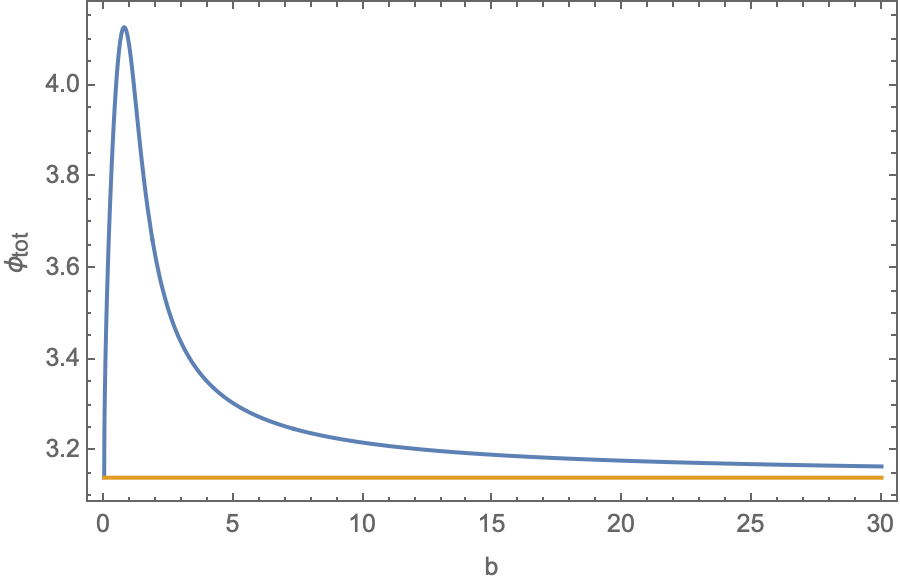} 
\includegraphics[width=0.32\textwidth]{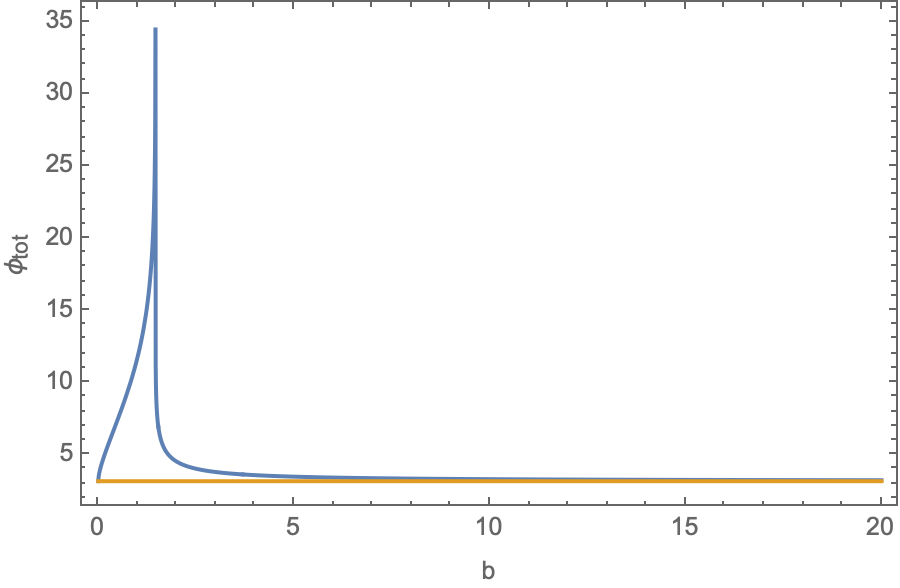} 
\caption{ \it Left plot: The emitted intensity as a function of $r$ for the naked singularity case. 
Middle and right plot: The total deflection angle as a function of the impact parameter $b$. 
The horizontal line in orange indicates $\phi=\pi$. 
The left plot is for the naked singularity without a light ring and the right one for the naked singularity with a light ring.}
\label{phitotns}
\end{figure}

A naked singularity can feature no light ring (type I) or one light ring (type II). 
Here we adopt the intensity function 
\begin{eqnarray}
&&I_{em}=
\begin{cases}
\frac{3}{(2r)^2}, & r\geq r_{ss}\\ \nn
0, &  r<r_{ss}.
\end{cases}
\label{intensityns}
\end{eqnarray}
We show $I_{em}$ in the left plot of Fig.~\ref{phitotns}.

\begin{figure}[t]
\centering
\includegraphics[width=0.4\textwidth]{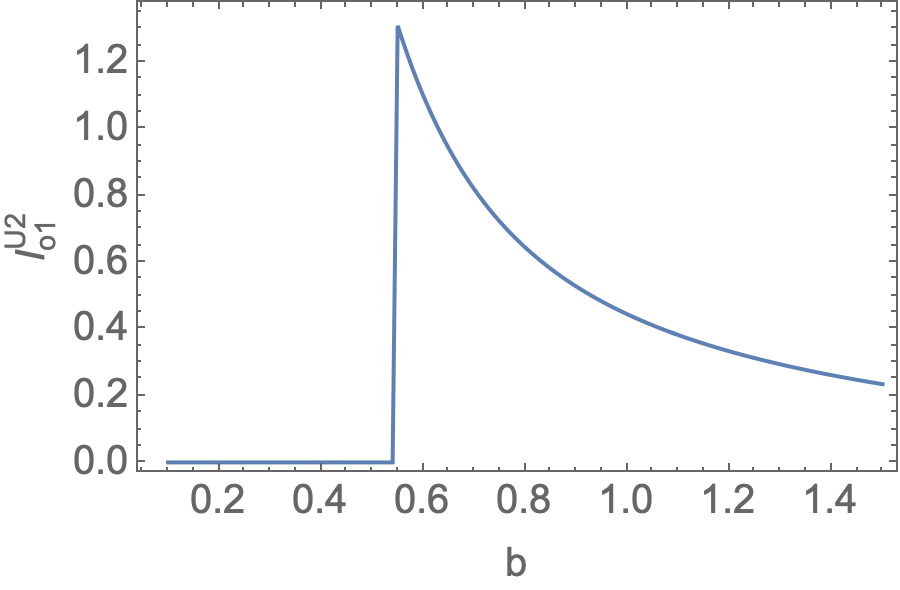} 
\includegraphics[width=0.42\textwidth]{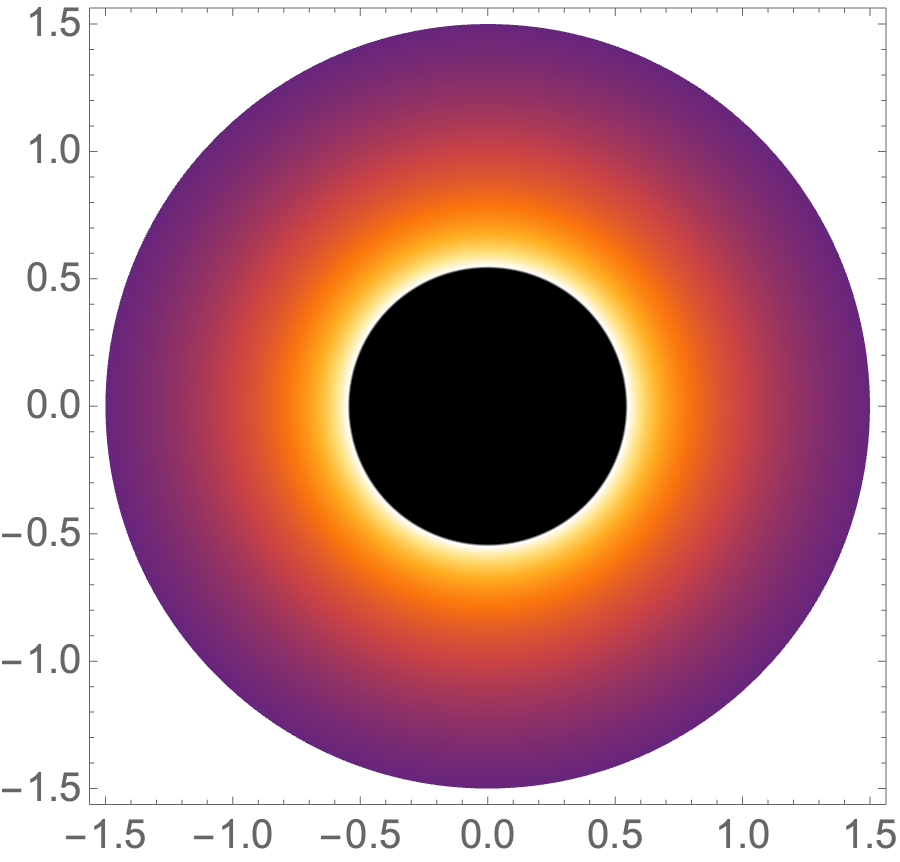} 
\caption{ \it The observed intensity and the corresponding optical image of the type I naked singularity without a light ring.}
\label{imagens1}
\end{figure}

For the type I naked singularity, the largest deflection angle of the light rays is less than $2\pi$. 
We present the deflection angle as a function of the impact parameter $b$ in the middle plot of Fig.~\ref{phitotns}. 
As for other naked singularities without a light ring, there are no bright rings seen in the images \cite{Joshi:2020tlq,Dey:2020bgo}. 
We present the deflection angle and an optical image in Fig.~\ref{imagens1}.
\begin{figure}[h]
\centering
\includegraphics[width=0.4\textwidth]{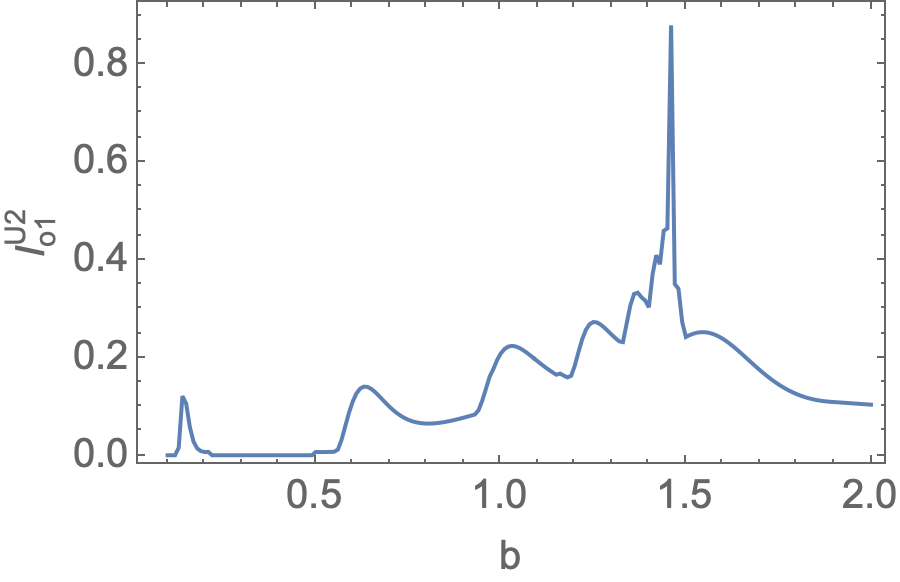} 
\includegraphics[width=0.42\textwidth]{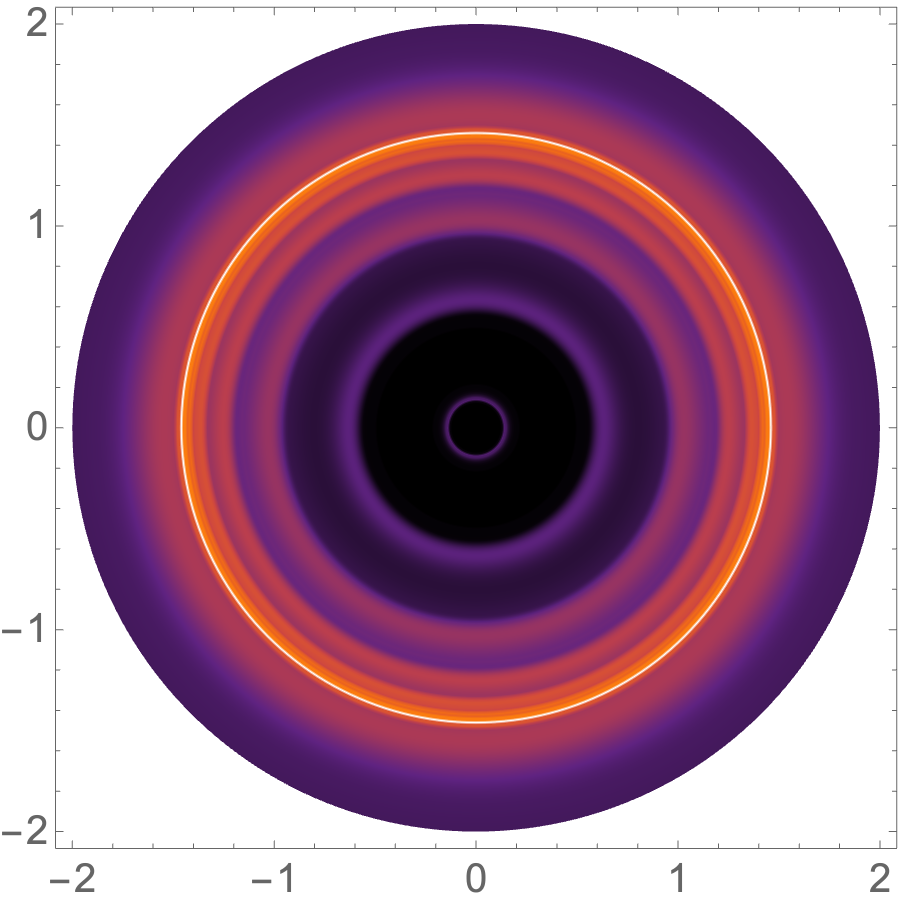} 
\caption{ \it The observed intensity and the corresponding optical image of the type II naked singularity with a light ring.}
\label{imagens2}
\end{figure}

For the type II naked singularity with a light ring, the total deflection angle tends to diverge at the light ring. 
We demonstrate this in the right plot of Fig.~\ref{phitotns}. 
Unlike the black hole case, the accretion disk extends from the static sphere located inside the light ring. 
In this case the optical image is very special. 
We present the deflection angle and an optical image in Fig.~\ref{imagens2}, which is qualitatively similar to the case of a wormhole with two unstable light rings in Ref.\cite{Guerrero:2022qkh}.

\newpage

\section{Conclusion}

The study of shadow images of compact objects is a valuable subject in both astrophysics and theoretical physics. 
Here we have focused on a beyond Horndeski theory, in which black holes, wormholes and naked singularities arise as solutions. 
We have investigated the timelike and null geodesics of these solutions. 

For the wormhole solutions we have found, that the innermost bound orbit is located at the wormhole throat, and they can possess one or two unstable light rings depending on the choice of the parameters.
The black hole solutions contain an ISCO outside the horizon and a light ring inside the ISCO. 
For the naked singularities we have found a static sphere, where massive particles with a suitable energy can stay at rest. 
The number of light rings of the naked singularities could be zero or one, but since the corresponding effective potential tends to infinity at the origin, no light rays can reach the naked singularity itself.

We have then considered optically and geometrically thin accretion disks illuminating the compact objects to produce optical images for wormholes, black holes, and naked singularities, respectively. 
Our results have shown that these objects possess quite different shadow images. 
Even though for the case when wormhole, black hole and naked singularity contain the same number of light rings (namely a single light ring), they 
exhibit different appearances. 
One reason is that the different location of the innermost bound orbit of massive particles affects the boundary of the associated accretion disk.

\section*{Acknowledgment}
We gratefully acknowledge support by DFG project Ku612/18-1 and the National Natural Science Foundation of China (NSFC) Grant No.~12205123 and Jiangxi Provincial Natural Science Foundation
with Grant No.~20232BAB211029 and by the Sino-German (CSC-DAAD) Postdoc Scholarship Program, No.~2021 (57575640).


\begin{thebibliography}{9}

\bibitem{Will:2005va}
C.~M.~Will,
``The Confrontation between general relativity and experiment,''
Living Rev. Rel. \textbf{9}, 3 (2006)

\bibitem{Will:2018bme}
C.~M.~Will,
``Theory and Experiment in Gravitational Physics",
(Cambridge University Press, 2018)

\bibitem{Berti:2015itd}
E.~Berti, E.~Barausse, V.~Cardoso, L.~Gualtieri, P.~Pani, U.~Sperhake, L.~C.~Stein, N.~Wex, K.~Yagi and T.~Baker, \textit{et al.}
``Testing General Relativity with Present and Future Astrophysical Observations,''
Class. Quant. Grav. \textbf{32}, 243001 (2015)

\bibitem{Barack:2018yly}
L.~Barack, V.~Cardoso, S.~Nissanke, T.~P.~Sotiriou, A.~Askar, C.~Belczynski, G.~Bertone, E.~Bon, D.~Blas and R.~Brito, \textit{et al.}
``Black holes, gravitational waves and fundamental physics: a roadmap,''
Class. Quant. Grav. \textbf{36}, 143001 (2019)

\bibitem{CANTATA:2021ktz}
E.~N.~Saridakis \textit{et al.} [CANTATA],
``Modified Gravity and Cosmology: An Update by the CANTATA Network,''
(Springer, 2021)

\bibitem{LIGOScientific:2016aoc}
B.~P.~Abbott \textit{et al.} [LIGO Scientific and Virgo],
``Observation of Gravitational Waves from a Binary Black Hole Merger,''
Phys. Rev. Lett. \textbf{116}, 061102 (2016)

\bibitem{LIGOScientific:2017vwq}
B.~P.~Abbott \textit{et al.} [LIGO Scientific and Virgo],
``GW170817: Observation of Gravitational Waves from a Binary Neutron Star Inspiral,''
Phys. Rev. Lett. \textbf{119}, 161101 (2017)

\bibitem{EventHorizonTelescope:2019dse}
K.~Akiyama \textit{et al.} [Event Horizon Telescope],
``First M87 Event Horizon Telescope Results. I. The Shadow of the Supermassive Black Hole,''
Astrophys. J. Lett. \textbf{875}, L1 (2019)

\bibitem{EventHorizonTelescope:2022wkp}
K.~Akiyama \textit{et al.} [Event Horizon Telescope],
``First Sagittarius A* Event Horizon Telescope Results. I. The Shadow of the Supermassive Black Hole in the Center of the Milky Way,''
Astrophys. J. Lett. \textbf{930}, L12 (2022)

\bibitem{Shao:2017gwu}
L.~Shao, N.~Sennett, A.~Buonanno, M.~Kramer and N.~Wex,
``Constraining nonperturbative strong-field effects in scalar-tensor gravity by combining pulsar timing and laser-interferometer gravitational-wave detectors,''
Phys. Rev. X \textbf{7}, 041025 (2017)

\bibitem{Freire:2022wcz}
P.~C.~C.~Freire,
``Tests of gravity theories with pulsar timing,''
[arXiv:2204.13468 [gr-qc]].

\bibitem{Morris:1988cz}
M.~S.~Morris and K.~S.~Thorne,
``Wormholes in space-time and their use for interstellar travel: A tool for teaching general relativity,''
Am. J. Phys. \textbf{56}, 395 (1988)

\bibitem{Morris:1988tu}
M.~S.~Morris, K.~S.~Thorne and U.~Yurtsever,
``Wormholes, Time Machines, and the Weak Energy Condition,''
Phys. Rev. Lett. \textbf{61}, 1446 (1988)


\bibitem{Cardoso:2016oxy}
V.~Cardoso, S.~Hopper, C.~F.~B.~Macedo, C.~Palenzuela and P.~Pani,
``Gravitational-wave signatures of exotic compact objects and of quantum corrections at the horizon scale,''
Phys. Rev. D \textbf{94}, 084031 (2016)


\bibitem{Mark:2017dnq}
Z.~Mark, A.~Zimmerman, S.~M.~Du and Y.~Chen,
``A recipe for echoes from exotic compact objects,''
Phys. Rev. D \textbf{96}, 084002 (2017)

\bibitem{Ou:2021efv}
M.~Y.~Ou, M.~Y.~Lai and H.~Huang,
``Echoes from asymmetric wormholes and black bounce,''
Eur. Phys. J. C \textbf{82}, 452 (2022)


\bibitem{Cunha:2022gde}
P.~Cunha, V.P., C.~Herdeiro, E.~Radu and N.~Sanchis-Gual,
``Exotic Compact Objects and the Fate of the Light-Ring Instability,''
Phys. Rev. Lett. \textbf{130}, 061401 (2023)


\bibitem{Horndeski:1974wa}
G.~W.~Horndeski,
``Second-order scalar-tensor field equations in a four-dimensional space,''
Int. J. Theor. Phys. \textbf{10}, 363 (1974)

\bibitem{Deffayet:2013lga}
C.~Deffayet and D.~A.~Steer,
``A formal introduction to Horndeski and Galileon theories and their generalizations,''
Class. Quant. Grav. \textbf{30}, 214006 (2013)

\bibitem{Kanti:1995vq}
P.~Kanti, N.~E.~Mavromatos, J.~Rizos, K.~Tamvakis and E.~Winstanley,
``Dilatonic black holes in higher curvature string gravity,''
Phys. Rev. D \textbf{54}, 5049 (1996)

\bibitem{Lu:2020iav}
H.~Lu and Y.~Pang,
``Horndeski gravity as $D \rightarrow 4$ limit of Gauss-Bonnet,''
Phys. Lett. B \textbf{809}, 135717 (2020)

\bibitem{Doneva:2022ewd}
D.~D.~Doneva, F.~M.~Ramazano\u{g}lu, H.~O.~Silva, T.~P.~Sotiriou and S.~S.~Yazadjiev,
``Scalarization,''
[arXiv:2211.01766 [gr-qc]].

\bibitem{Kanti:2011jz}
P.~Kanti, B.~Kleihaus and J.~Kunz,
``Wormholes in Dilatonic Einstein-Gauss-Bonnet Theory,''
Phys. Rev. Lett. \textbf{107}, 271101 (2011)

\bibitem{Antoniou:2019awm}
G.~Antoniou, A.~Bakopoulos, P.~Kanti, B.~Kleihaus and J.~Kunz,
``Novel Einstein\textendash{}scalar-Gauss-Bonnet wormholes without exotic matter,''
Phys. Rev. D \textbf{101}, 024033 (2020)

\bibitem{Kleihaus:2019rbg}
B.~Kleihaus, J.~Kunz and P.~Kanti,
``Particle-like ultracompact objects in Einstein-scalar-Gauss-Bonnet theories,''
Phys. Lett. B \textbf{804}, 135401 (2020)



\bibitem{Glavan:2019inb}
D.~Glavan and C.~Lin,
``Einstein-Gauss-Bonnet Gravity in Four-Dimensional Spacetime,''
Phys. Rev. Lett. \textbf{124}, 081301 (2020)

\bibitem{Hennigar:2020lsl}
R.~A.~Hennigar, D.~Kubiz\v{n}\'ak, R.~B.~Mann and C.~Pollack,
``On taking the D \textrightarrow{} 4 limit of Gauss-Bonnet gravity: theory and solutions,''
JHEP \textbf{07}, 027 (2020)

\bibitem{Fernandes:2020nbq}
P.~G.~S.~Fernandes, P.~Carrilho, T.~Clifton and D.~J.~Mulryne,
``Derivation of Regularized Field Equations for the Einstein-Gauss-Bonnet Theory in Four Dimensions,''
Phys. Rev. D \textbf{102}, 024025 (2020)

\bibitem{Gleyzes:2014dya}
J.~Gleyzes, D.~Langlois, F.~Piazza and F.~Vernizzi,
``Healthy theories beyond Horndeski,''
Phys. Rev. Lett. \textbf{114}, 211101 (2015)

\bibitem{Langlois:2015cwa}
D.~Langlois and K.~Noui,
``Degenerate higher derivative theories beyond Horndeski: evading the Ostrogradski instability,''
JCAP \textbf{02}, 034 (2016)


\bibitem{Gleyzes:2014qga}
J.~Gleyzes, D.~Langlois, F.~Piazza and F.~Vernizzi,
``Exploring gravitational theories beyond Horndeski,''
JCAP \textbf{02}, 018 (2015)

\bibitem{Crisostomi:2016czh}
M.~Crisostomi, K.~Koyama and G.~Tasinato,
``Extended Scalar-Tensor Theories of Gravity,''
JCAP \textbf{04}, 044 (2016)




\bibitem{Crisostomi:2016tcp}
M.~Crisostomi, M.~Hull, K.~Koyama and G.~Tasinato,
``Horndeski: beyond, or not beyond?,''
JCAP \textbf{03}, 038 (2016)



\bibitem{Bambi:2012tg}
C.~Bambi,
``A code to compute the emission of thin accretion disks in non-Kerr space-times and test the nature of black hole candidates,''
Astrophys. J. \textbf{761}, 174 (2012)

\bibitem{Johannsen:2013vgc}
T.~Johannsen,
``Photon Rings around Kerr and Kerr-like Black Holes,''
Astrophys. J. \textbf{777}, 170 (2013)



\bibitem{Cunha:2015yba}
P.~V.~P.~Cunha, C.~A.~R.~Herdeiro, E.~Radu and H.~F.~Runarsson,
``Shadows of Kerr black holes with scalar hair,''
Phys. Rev. Lett. \textbf{115}, 211102 (2015)


\bibitem{Cunha:2016bpi}
P.~V.~P.~Cunha, C.~A.~R.~Herdeiro, E.~Radu and H.~F.~Runarsson,
``Shadows of Kerr black holes with and without scalar hair,''
Int. J. Mod. Phys. D \textbf{25}, 1641021 (2016)

\bibitem{Gralla:2019xty}
S.~E.~Gralla, D.~E.~Holz and R.~M.~Wald,
``Black Hole Shadows, Photon Rings, and Lensing Rings,''
Phys. Rev. D \textbf{100}, 024018 (2019)


\bibitem{Younsi:2021dxe}
Z.~Younsi, D.~Psaltis and F.~\"Ozel,
``Black Hole Images as Tests of General Relativity: Effects of Spacetime Geometry,''
Astrophys. J. \textbf{942}, 47 (2023)

\bibitem{Promsiri:2023rez}
C.~Promsiri, W.~Horinouchi and E.~Hirunsirisawat,
``Remarks on the light ring images and the optical appearance of hairy black holes in Einstein-Maxwell-dilaton gravity,''
[arXiv:2310.04221 [gr-qc]].





\bibitem{Kumar:2020owy}
R.~Kumar and S.~G.~Ghosh,
``Rotating black holes in $4D$ Einstein-Gauss-Bonnet gravity and its shadow,''
JCAP \textbf{07}, 053 (2020)


\bibitem{Guo:2020zmf}
M.~Guo and P.~C.~Li,
``Innermost stable circular orbit and shadow of the $4D$ Einstein\textendash{}Gauss\textendash{}Bonnet black hole,''
Eur. Phys. J. C \textbf{80}, 588 (2020)



\bibitem{Konoplya:2020bxa}
R.~A.~Konoplya and A.~F.~Zinhailo,
``Quasinormal modes, stability and shadows of a black hole in the 4D Einstein\textendash{}Gauss\textendash{}Bonnet gravity,''
Eur. Phys. J. C \textbf{80}, 1049 (2020)


\bibitem{Zeng:2020dco}
X.~X.~Zeng, H.~Q.~Zhang and H.~Zhang,
``Shadows and photon spheres with spherical accretions in the four-dimensional Gauss\textendash{}Bonnet black hole,''
Eur. Phys. J. C \textbf{80}, 872 (2020)

\bibitem{Peng:2020wun}
J.~Peng, M.~Guo and X.~H.~Feng,
``Influence of quantum correction on black hole shadows, photon rings, and lensing rings,''
Chin. Phys. C \textbf{45}, 085103 (2021)

\bibitem{Gyulchev:2021dvt}
G.~Gyulchev, P.~Nedkova, T.~Vetsov and S.~Yazadjiev,
``Image of the thin accretion disk around compact objects in the Einstein\textendash{}Gauss\textendash{}Bonnet gravity,''
Eur. Phys. J. C \textbf{81}, 885 (2021)


\bibitem{Zeng:2022fdm}
X.~X.~Zeng, M.~I.~Aslam and R.~Saleem,
``The optical appearance of charged four-dimensional Gauss\textendash{}Bonnet black hole with strings cloud and non-commutative geometry surrounded by various accretions profiles,''
Eur. Phys. J. C \textbf{83}, 129 (2023)


\bibitem{Ye:2023qks}
J.~P.~Ye, Z.~Q.~He, A.~X.~Zhou, Z.~Y.~Huang and J.~H.~Huang,
``Shadows and photon rings of a quantum black hole,''
[arXiv:2312.17724 [gr-qc]].





\bibitem{Bambi:2013nla}
C.~Bambi,
``Can the supermassive objects at the centers of galaxies be traversable wormholes? The first test of strong gravity for mm/sub-mm very long baseline interferometry facilities,''
Phys. Rev. D \textbf{87}, 107501 (2013)

\bibitem{Tsukamoto:2021fpp}
N.~Tsukamoto,
``Linearization stability of reflection-asymmetric thin-shell wormholes with double shadows,''
Phys. Rev. D \textbf{103}, 064031 (2021)

\bibitem{Guerrero:2021pxt}
M.~Guerrero, G.~J.~Olmo and D.~Rubiera-Garcia,
``Double shadows of reflection-asymmetric wormholes supported by positive energy thin-shells,''
JCAP \textbf{04}, 066 (2021)


\bibitem{Peng:2021osd}
J.~Peng, M.~Guo and X.~H.~Feng,
``Observational signature and additional photon rings of an asymmetric thin-shell wormhole,''
Phys. Rev. D \textbf{104}, 124010 (2021)

\bibitem{Bambi:2021qfo}
C.~Bambi and D.~Stojkovic,
``Astrophysical Wormholes,''
Universe \textbf{7}, 136 (2021)

\bibitem{Rahaman:2021web}
F.~Rahaman, K.~N.~Singh, R.~Shaikh, T.~Manna and S.~Aktar,
``Shadows of Lorentzian traversable wormholes,''
Class. Quant. Grav. \textbf{38}, 215007 (2021)


\bibitem{Schee:2021pdt}
J.~Schee and Z.~Stuchl\'\i{}k,
``Appearance of Keplerian discs orbiting on both sides of reflection-symmetric wormholes,''
JCAP \textbf{01}, 054 (2022)

\bibitem{Guerrero:2022qkh}
M.~Guerrero, G.~J.~Olmo, D.~Rubiera-Garcia and D.~G\'omez S\'aez-Chill\'on,
``Light ring images of double photon spheres in black hole and wormhole spacetimes,''
Phys. Rev. D \textbf{105}, 084057 (2022)

\bibitem{Delijski:2022jjj}
V.~Delijski, G.~Gyulchev, P.~Nedkova and S.~Yazadjiev,
``Polarized image of equatorial emission in horizonless spacetimes: Traversable wormholes,''
Phys. Rev. D \textbf{106}, 104024 (2022)


\bibitem{Huang:2023yqd}
H.~Huang, J.~Kunz, J.~Yang and C.~Zhang,
``Light ring behind wormhole throat: Geodesics, images, and shadows,''
Phys. Rev. D \textbf{107}, 104060 (2023)

\bibitem{Ishkaeva:2023xny}
V.~A.~Ishkaeva and S.~V.~Sushkov,
``Image of an accreting general Ellis-Bronnikov wormhole,''
Phys. Rev. D \textbf{108}, 084054 (2023)









\bibitem{Shaikh:2018lcc}
R.~Shaikh, P.~Kocherlakota, R.~Narayan and P.~S.~Joshi,
``Shadows of spherically symmetric black holes and naked singularities,''
Mon. Not. Roy. Astron. Soc. \textbf{482}, 52 (2019)

\bibitem{Gyulchev:2019tvk}
G.~Gyulchev, P.~Nedkova, T.~Vetsov and S.~Yazadjiev,
``Image of the Janis-Newman-Winicour naked singularity with a thin accretion disk,''
Phys. Rev. D \textbf{100}, 024055 (2019)

\bibitem{Gyulchev:2020cvo}
G.~Gyulchev, J.~Kunz, P.~Nedkova, T.~Vetsov and S.~Yazadjiev,
``Observational signatures of strongly naked singularities: image of the thin accretion disk,''
Eur. Phys. J. C \textbf{80}, 1017 (2020)

\bibitem{Joshi:2020tlq}
A.~B.~Joshi, D.~Dey, P.~S.~Joshi and P.~Bambhaniya,
``Shadow of a Naked Singularity without Photon Sphere,''
Phys. Rev. D \textbf{102}, 024022 (2020)



\bibitem{Dey:2020bgo}
D.~Dey, R.~Shaikh and P.~S.~Joshi,
``Shadow of nulllike and timelike naked singularities without photon spheres,''
Phys. Rev. D \textbf{103}, 024015 (2021)


\bibitem{Vagnozzi:2022moj}
S.~Vagnozzi, R.~Roy, Y.~D.~Tsai, L.~Visinelli, M.~Afrin, A.~Allahyari, P.~Bambhaniya, D.~Dey, S.~G.~Ghosh and P.~S.~Joshi, \textit{et al.}
``Horizon-scale tests of gravity theories and fundamental physics from the Event Horizon Telescope image of Sagittarius A,''
Class. Quant. Grav. \textbf{40}, 165007 (2023)

\bibitem{Tavlayan:2023vbv}
A.~Tavlayan and B.~Tekin,
``Instability of a Kerr-type naked singularity due to light and matter accretion and its shadow,''
[arXiv:2301.13751 [gr-qc]].

\bibitem{Deliyski:2023gik}
V.~Deliyski, G.~Gyulchev, P.~Nedkova and S.~Yazadjiev,
``Polarized image of equatorial emission in horizonless spacetimes: Naked singularities,''
Phys. Rev. D \textbf{108}, no.10, 104049 (2023)
doi:10.1103/PhysRevD.108.104049
[arXiv:2303.14756 [gr-qc]].



\bibitem{Rosa:2022tfv}
J.~L.~Rosa and D.~Rubiera-Garcia,
``Shadows of boson and Proca stars with thin accretion disks,''
Phys. Rev. D \textbf{106}, 084004 (2022)



\bibitem{Rosa:2022toh}
J.~L.~Rosa, P.~Garcia, F.~H.~Vincent and V.~Cardoso,
``Observational signatures of hot spots orbiting horizonless objects,''
Phys. Rev. D \textbf{106}, 044031 (2022)






\bibitem{Bakopoulos:2021liw}
A.~Bakopoulos, C.~Charmousis and P.~Kanti,
``Traversable wormholes in beyond Horndeski theories,''
JCAP \textbf{05}, 022 (2022)

\bibitem{BenAchour:2016fzp}
J.~Ben Achour, M.~Crisostomi, K.~Koyama, D.~Langlois, K.~Noui and G.~Tasinato,
``Degenerate higher order scalar-tensor theories beyond Horndeski up to cubic order,''
JHEP \textbf{12}, 100 (2016)

\bibitem{Kobayashi:2011nu}
T.~Kobayashi, M.~Yamaguchi and J.~Yokoyama,
``Generalized G-inflation: Inflation with the most general second-order field equations,''
Prog. Theor. Phys. \textbf{126}, 511 (2011)

\bibitem{Grandclement:2014msa}
P.~Grandclement, C.~Som\'e and E.~Gourgoulhon,
``Models of rotating boson stars and geodesics around them: new type of orbits,''
Phys. Rev. D \textbf{90}, 024068 (2014)


\bibitem{Teodoro:2020kok}
M.~C.~Teodoro, L.~G.~Collodel and J.~Kunz,
``Retrograde Polish Doughnuts around Boson Stars,''
JCAP \textbf{03}, 063 (2021)



\bibitem{Gao:2023mjb}
X.~J.~Gao, T.~T.~Sui, X.~X.~Zeng, Y.~S.~An and Y.~P.~Hu,
``Investigating shadow images and rings of the charged Horndeski black hole illuminated by various thin accretions,''
Eur. Phys. J. C \textbf{83}, 1052 (2023)


\bibitem{Wei:2023bgp}
S.~W.~Wei, Y.~P.~Zhang, Y.~X.~Liu and R.~B.~Mann,
``Static spheres around spherically symmetric black hole spacetime,''
Phys. Rev. Res. \textbf{5}, 043050 (2023)



\end{thebibliography}
\end{document}